\documentclass[11pt,preprint]{aastex}
\def\fre#1{\frac{ {\hbox{$ #1 $}} } }
\def\fri#1{ {\hbox{$ #1 $}}  }
\begin{document}
\title{Line-driven winds, ionizing fluxes and UV-spectra of hot stars at
      extremely low metallicity. I. Very massive O-stars}
\author{Rolf P. Kudritzki}
\affil{Institute for Astronomy, University of Hawaii, 2680 Woodlawn Drive,
       Honolulu, HI 96822}
\email{kudritzki@usm.uni-muenchen.de}


\begin{abstract}
Wind models of very massive stars with metallicities in a range
from $10^{-4}$  to 1.0 solar are calculated using a new treatment
of radiation driven winds with depth dependent radiative force
multipliers and a comprehensive list of more than two million of
spectral lines in NLTE. The models are tested by a comparison with
observed stellar wind properties of O stars in the Galaxy and the
SMC. Satisfying agreement is found. The calculations yield
mass-loss rates, wind velocities, wind momenta and wind energies
as a function of metallicity and can be used to discuss the
influence of stellar winds on the evolution of very massive stars
in the early universe and on the interstellar medium in the early
phases of galaxy formation. It is shown that the normal scaling
laws, which predict stellar mass-loss rates and wind momenta to
decrease as a power law with metal abundance break down at a
certain threshold. Analytical fit formulae for mass-loss rates are
provided as a function of stellar parameters and metallicity.

Ionizing fluxes of hot stars depend crucially on the strengths of
their stellar winds, which modify the absorption edges of hydrogen
and helium (neutral and ionized) and the line blocking in the far
UV. The new wind models are therefore also applied to calculate
ionizing fluxes and observable spectra of very massive stars as a
function of metallicity using the new hydrodynamic, non-LTE
line-blanketed flux constant model atmosphere code developed by
\citet{pau01}. Numbers of ionizing photons for the
crucial ionization stages are given. For a fixed effective
temperature the He II ionizing emergent flux depends very strongly
on metallicity but also on stellar luminosity. A strong dependence
on metallicity is also found for the C III, Ne II and O II
ionizing photons, whereas the H I and He I ionizing flux is almost
independent of metallicity. We also calculate UV spectra for all
the models and discuss the behaviour of significant line features
as a function of metallicity.

\end{abstract}


\section{Introduction}

There is growing evidence that the evolution of galaxies in the
early universe was heavily influenced by the formation of first
generations of very massive stars. Numerical simulations indicate
that for metallicities  $Z/Z_{\odot}$ of the order of $10^{-3}$ or
smaller the formation of stars with masses larger than 100
M$_{\odot}$ is strongly favored and that the Initial Mass Function
becomes top-heavy and deviates significantly from the standard
(Salpeter) power-law \cite{abel00}, \cite{abel02}, \cite{bro99}, \cite{bro01a},
\cite{bro02a}, \cite{nak01}. Such an early
population of very massive stars at very low metallicities could
contribute significantly to the ionization history of the
intergalactic medium (\citealt{car84}, \citealt{cou86},
\citealt{hai97}, \citealt{bro01b}), which appears to
have been reionized at redshifts z possibly larger than 6, as the recent
work on quasars (\citealt{bec01}, \citealt{djo01a}, \citealt{fan00}, \citealt{fan02})
and Ly$_{\alpha}$-emitters (\citealt{hu02}) indicates. Very massive
stars are also very likely
the progenitors of Gamma-Ray bursts (\citealt{mac01},
\citealt{djo01b}, \citealt{kul00}, \citealt{rei01}),
which then  - as tracers of the cosmic star formation history - might
originate to a large fraction at very high reshift (\citealt{bro02b},
\citealt{cia00}, \citealt{lam00}). In
addition, the extreme Lyman $\alpha$ emitting galaxies at high red
redshift can be explained by an ionizing population of very
massive stars at very low metallicity (\citealt{kud00},
\citealt{rho02}, \citealt{mal02}).

In order to be able to make more quantitative predictions about
the influence of such an extremely metal-poor population of very
massive stars on their galactic and intergalactic environment, one
needs to determine their physical properties during their
evolution. A key issue in this regard is the knowledge about their
stellar winds.

All hot massive stars have winds driven by radiation. These winds
have substantial effects on the structure of the radiating
atmospheres. They dominate the density stratification and the
radiative transfer through the presence of their macroscopic
transonic velocity fields (see \citealt{kud98} for a detailed
review) and they modify the amount of the emergent ionizing
radiation significantly (\citealt{gab89}, \citealt{gab91}, \citealt{gab92},
\citealt{naj96}). Winds have an extremely important influence
on the evolution of massive stars by reducing the stellar mass
continuously and by affecting evolutionary time-scales, chemical
profiles, surface abundances and luminosities. Providing a
significant input of mechanical and radiative momentum and energy
into the ISM together with the injection of nuclear processed
material they can also play a crucial role for the evolution of
galaxies. Last but not least, stellar winds provide beautiful
spectroscopic tools to investigate the physical properties of
galaxies through the analysis of broad stellar wind spectral line
features easily detectable in the integrated spectra of
starforming galaxies in the nearby and high-redshift universe
(\citealt{pet00}, \citealt{lei01}).

While the investigation of winds from massive stars in the solar
neighborhood and the Magellanic Clouds has been the subject of
extensive work (see \citealt{kud00}, for a recent review), little is
known so far about winds at very low metallicity. Since these
winds are initiated and maintained through absorption of
photospheric photon momentum by UV metal lines, we expect their
strengths to decrease with decreasing metallicity. \cite{kud87}
were the first to calculate radiation driven wind models in a
metallicity range from solar to 0.1 solar and predicted that
mass-loss rates scale with $(Z/Z_{\odot})^{0.5 .. 0.8}$.
\cite{lei92}) confirmed this conclusion by independent
calculations, which were extended to include a few models with
metallicities as low as 0.01 solar. Very recently, \cite{vin01}
provided new wind models for normal O-stars and B-supergiants and
obtained again a very similar power law. Observational
spectroscopic studies of massive stars in the Magellanic Clouds
confirm that this theoretical prediction is basically correct for
metallicities ranging to 0.2 solar (see \citealt{kud00}, for a
review, but also \citealt{vin01}).

The objective of the work presented here is to study the mechanism of
radiative line driving and the corresponding properties of the winds of
possible generations of very massive stars at extremely low
metallicities and to investigate the principal influence of these
winds on ionizing fluxes and observable ultraviolet spectra. As we
will demonstrate in sections 2 and 3, the very low
metallicities require the development of a new approach to
calculate the wind dynamics. The basic new element of this
approach, needed in the domain of extremely low metallicity, is
the introduction of depth dependent force multipliers representing
the radiative line acceleration. To calculate our wind models we
take into account the improvements accomplished during the last
decade with regard to atomic physics and line lists (see \citealt{pau98},
\citealt{pau01}). We use the line list of 2.5 10$^{6}$ lines of
150 ionic species and apply analytical formulae (see \citealt{luc93},
\citealt{spr97}, \citealt{spr98}, \citealt{pul00}) for a fast
approximation of NLTE occupation numbers to
calculate the radiative line acceleration, which is then
represented by the above mentioned new parameterization using
depth dependent force multiplier parameters. Because of the depth
dependent force multipliers a new formulation of the critical
point equations is developed and a new iterative solution
algorithm for the complete stellar wind problem is introduced
(section 4). This new approach includes the old algorithm
\cite{pau86} in the limit of force multipliers, which
do not depend on depth. It allows to calculate wind models within a
few seconds on a workstation for every hot star with specified
effective temperature, mass, radius and abundances.

In section 5 we will test our new algorithm by comparing with
observed wind properties for galactic and Magellanic Cloud
O-stars. In section 6 we will then extent these calculations to
significantly higher masses and luminosities and present and
discuss wind models in a metallicity range down to $10^{-4}$
solar. In section 7 we will discuss spectral energy distributions,
numbers of ionizing photons and line spectra. We will conclude
with a short discussion and the perspectives of future work in
section 8.

Our investigation concentrates on mass-loss through radiation driven winds
only. As is well
known, very massive stars are pulsationally unstable, which might contribute
to stellar-mass loss, in particular at low metallicity when the contribution
of the radiative driving to the winds decreases. However,
very recently, \cite{bar01} have studied this problem and found
that the possible effects of pulsation on mass-loss are much weaker for
very massive stars with low
metallicity than for those with solar metallicity. It is thus very
likely that the mechanism of radiative line driving remains still important
in the metallicity range discussed here, although pulsational instability
will probably lead to an additional mass-loss contribution for our lowest
metallicity models.

\section{Radiative acceleration and effective gravity at very
low metallicities}

The hydrodynamics of stationary and spherical symmetric line driven winds
are described by the equation of continuity

\begin{equation}
\dot{M} = 4\pi r^{2} \rho(r) v(r)
\label{econt}
\end{equation}

and the equation of motion

\begin{equation}
v(r) \fre{dv(r)}{\fri{dr}}=-\fre{1}{\fri{\rho(r)}} \fre{dP_{gas}(r)}{\fri{dr}}
- g_{eff}(r).
\label{eom1}
\end{equation}

Here $v(r)$ is the the velocity field as a function of the radial
coordinate r, $\rho(r)$ is the density distribution, $P(r)_{gas}$
is the gas pressure and $\dot{M}$ is the mass-los rate. The
effective gravity $g_{eff}(r)$ is the difference between the
gravitational and radiative acceleration.

\begin{equation}
g_{eff}(r) = g(r) - g_{rad}(r).
\label{geff}
\end{equation}

The radiative acceleration consists of three terms

\begin{equation}
g_{rad}(r) = g^{Th}_{rad} + g^{bf,ff}_{rad} + g^{lines}_{rad}
\label{grad}
\end{equation}

representing the contributions of Thomson scattering by free electrons
($g^{Th}_{rad}$), bound-free and free-free absorption ($g^{bf,ff}_{rad}$)
and line absorption ($g^{lines}_{rad}$), respectively. In the outer
atmospheric layers of
hot stars, where winds start to become significant, the particle densities
$n_{e}$ and $n_{p}$
of electrons and protons are usually smaller than 10$^{12.5}$cm$^{-3}$ (except
for Wolf-Rayet stars and very extreme supergiants) so that the contribution
of $g^{bf,ff}_{rad}$ can be neglected.

The crucial term in the hydrodynamics of radiation driven winds is the
radiative line acceleration, which can be expressed in
units of the Thomson acceleration

\begin{equation}
g^{lines}_{rad} = g^{Th}_{rad} CF(r,v,\fre{dv}{\fri{dr}}) {\it M}(t).
\label{gradl}
\end{equation}

$CF(r,v,\fre{dv}{\fri{dr}})$ is the finite cone angle correction factor, which
takes into account that a volume element in the stellar wind is irradiated by
a stellar disk of finite angular diameter rather than a point source. (For a
discussion of CF see \citealt{pau86}, and \citealt{kud89}).
{\it M}(t) is the {\it line force multiplier} which gives the line
acceleration in
units of Thomson scattering. In the Sobolev approximation the contribution of
all spectral lines $i$ at frequencies $\nu_{i}$ and at spectral luminosities
$L_{\nu_{i}}$ to the line force multiplier is given by

\begin{equation}
{\it M}(t) = \fre{v_{therm}}{\fri{c}} \fre{1}{\fri{t}}
\sum_{i} \fre{\nu_{i}L_{\nu_{i}}} {\fri{L}}
(1-e^{-\tau_{i}}),
\label{fmpexact}
\end{equation}

where $v_{therm}$ is the thermal velocity of hydrogen, $c$ the speed of
light and

\begin{equation}
\tau_{i} = k_{i} t(r)
\label{tau}
\end{equation}

is the local (Sobolev) optical depth of line $i$ computed as product of two
factors, the line strength $k_{i}$

\begin{equation}
k_{i} \propto \fre{n_{l}}{\fri{n_{e}}} f_{lu} \lambda_{i}
\label{kline}
\end{equation}

and the Thomson optical depth parameter $t(r)$

\begin{equation}
t(r) = n_{e}\sigma_{e} \fre{v_{therm}}{\fri{dv/dr}}
\label{tpara}
\end{equation}

(for details, see \citealt{cas75}, \citealt{abb82}, \citealt{pau86},
\citealt{kud88} and \citealt{kud98} and references therein).

Eqs.\,\ref{fmpexact}, \ref{tau}, \ref{kline} and \ref{tpara} allow already
a first discussion what to expect for line driven winds at extremely low
metallicities. In such a situation we expect very weak winds of low
density leading to very small optical thickness parameters t. In the most
extreme case t could become so small that even for the lines with the
strongest line strengths the Sobolev optical depth would be smaller than
unity. Then, the line force multiplier would become independent of t and
saturate at its maximum value

\begin{equation}
{\it M}^{max} = \fre{v_{therm}}{\fri{c}}
\sum_{i} \fre{\nu_{i}L_{\nu_{i}}} {\fri{L}} k_{i}.
\label{fmpmax1}
\end{equation}

A typical value for O-stars with solar metallicity is ${\it M}^{max}$ = 2000
(\citealt{gay95}, \citealt{pul00}). On the other hand, it is certainly
reasonable to assume that the line strengths k$_{i}$ of the individual metal
lines are proportional to the metallicity Z

\begin{equation}
k_{i} = k^{\odot}_{i} \fre{Z}{\fri{Z_{\odot}}}
\label{klinemet}
\end{equation}

resulting in a simple zero-order estimate for the maximum line force
multiplier as a function of metallicity (${\it M}^{H,He}$ is the contribution
of the hydrogen and helium lines to the force multiplier)

\begin{equation}
{\it M}^{max} = 2000 \fre{Z}{\fri{Z_{\odot}}} + {\it M}^{H,He}.
\label{fmpmet}
\end{equation}

The existence of a line driven wind requires as a necessary condition
that the effective gravity has to become negative somewhere.
Using Eqs.\,\ref{geff}, \ref{grad}, \ref{gradl} (and CF=1, for simplicity)
we derive

\begin{equation}
\Gamma \geq \Gamma_{min} = \fre{1}{\fri{1+ {\it M}^{max}}} =
\lbrace 1 + 2000 \fre{Z}{\fri{Z_{\odot}}} + {\it M}^{H,He} \rbrace ^{-1},
\label{gammamin}
\end{equation}

where $\Gamma$ is the usual distance to the Eddington-limit

\begin{equation}
\Gamma = g^{Th}_{rad}(r)/g(r).
\label{gamma}
\end{equation}

Assuming that the contribution by {\it M}$^{H,He}$ is very small (but see
section 3) we conclude
from Eq.\,\ref{gammamin} that stars at extremely low metallicity
can only have line driven winds, if they are very close to the
Eddington limit. For Z/Z$_{\odot} = 10^{-2}$ wind solutions are still possible
in a wide range,
as we obtain $\Gamma_{min} = 0.05$. However, for Z/Z$_{\odot} = 10^{-3}$ and
$10^{-4}$ we find $\Gamma_{min}$ = 1/3 and 5/6, respectively, indicating
where to expect winds, if the metallicity is extremely low.

\section{A new parametrization of the radiative line force}

In a realistic hydrodynamic stellar-wind code the contributions of hundred
thousands of lines are added up to calculate the radiative acceleration at
every depth point and to solve for the velocity field and the mass-loss rate
(\citealt{pau94}). However , for computational reasons these numbers are
not used directly to solve the hydrodynamical problem. Instead, following the
pioneering work of \cite{cas75} and \cite{abb82}, ${\it M}(t)$ is
usually fitted by the parametrization ($W(r)$ is the geometrical dilution
factor of the radiation field)

\begin{equation}
{M}(t) = \hat{k} t^{-\alpha} \hat{n}^{\delta}, \qquad \qquad
\hat{n} = \fre{n_{e}(r)}{\fri{W(r)}}/10^{11}cm^{-3}.
\label{lfm1}
\end{equation}

{\bf $\hat{k}$, $\alpha$, $\delta$} are the force multiplier parameters
(fmps). This
parametrization has the advantage that it allows very fast numerical solutions
(see \citealt{pau86} and very precise analytical approximations
of the complex hydrodynamical problem of line driven winds
(see \citealt{kud89}), if one assumes that the fmps are constant in the
atmosphere.

The idea behind this parametrization (for a more detailed discussion
see, for instance \citealt{owo88}, \citealt{kud98}, \citealt{gay95} and \citealt{pul00})
is that to some
approximation the distribution function of line strengths $n(k,\nu)$
obeys a power law

\begin{equation}
n(k,\nu)d\nu dk = (1-\alpha)g(\nu)d\nu k^{{\bf \alpha}-2} dk,
\label{nk}
\end{equation}

at all frequencies $\nu$. The exponent $\alpha$, which physically
describes the steepness of the line strengths distribution function,
is mostly determined by the atomic physics of the dominant ionization stages
and basically reflects the distribution function of the oscillator strengths.
Typical values vary between

\begin{equation}
\alpha = 0.5 \ldots 0.7.
\end{equation}

If the sum in ${\it M}(t)$ is replaced by a double integral (in
frequency and line strength) assuming that the line strength
distribution function follows Eq.\,\ref{nk} over the full range of
line-strengths from zero to infinity, then the first two factors
of Eq.\,\ref{lfm1}  are obtained. The fmp $\hat{k}$ then relates
to

\begin{equation}
N_{eff} = \int^{\infty}_{0} \fre{\nu L_{\nu}}{\fri{L}} g(\nu) d\nu
\label{lnorm}
\end{equation}

the frequency normalization of the line strength distribution
function. Since $N_{eff}$ changes, if the ionization changes in
the stellar wind, and since in NLTE the ionization balance to
first order is determined by the ratio of electron density $n_{e}$
to geometrical dilution $W(r)$ of the radiation field, the third
factor is introduced. Typical values of $\delta$ for hot stars of
solar metallicity are in a range of

\begin{equation}
\delta = 0.0 \ldots 0.2.
\end{equation}

It is important to note that neglecting the
ionization dependence of the force multiplier described by the third factor
leads to unrealistic stellar wind stratifications, as soon as even mild
changes in the ionization become important. On the other hand, we also realize
that accounting for ionization effects in the form of Eq.\,\ref{lfm1}
assumes that
the exponent $\alpha$ of the line strength distribution function does not
depend on ionization. We will see later that this is not true.

After the discussion in section 2 it is evident that the parametrization of
Eq.\,\ref{lfm1} can not be valid over an unlimited range of optical depth
parameters t. The reason is that in reality there are no lines with infinite
line strengths. There will always be a line with maximum line strength
$k_{max}$ and, consequently, an optical depth parameter t$_{sat}$ below which
all lines contributing to the radiative acceleration are optically thin so
that the line force multiplier saturates at ${\it M}$(t) = ${\it M}$$^{max}$
for t/t$_{sat} \le $ 1. For 1 $ \le$ t/t$_{sat} \le $ 10$^{2}$  the
t-dependence of ${\it M}$(t) becomes flatter and, if fitted by a power law,
the local exponent defined as

\begin{equation}
\alpha(t,\hat{n}) = -\fre{\partial log({\it M}(t,\hat{n}))}{\fri{\partial
log(t)}}
\label{alpha}
\end{equation}

becomes smaller. For low metallicities this effect becomes more important,
since according to Eq.\,\ref{klinemet} t$_{sat}$ will shift
to larger values of t
and, in addition, the winds will become weaker resulting in
smaller optical depth parameters throughout the wind.

The cutoff of line strengths at the high end is not the only important effect.
In addition, there are systematic deviations from a strict power law resulting
in a curvature of $log (n(k,\nu))$ if plotted against $log (k)$ (see
\citealt{kud98}). This effect and the atomic and
statistic physics behind it have been
carefully and extensively investigated by \cite{pul00}. (For a deeper
understanding of the physics of the line strength distribution function we
refer the reader to their paper). For O-star temperatures, the slope
$\fre{\partial log (n(k,\nu))}{\fri{\partial log(k)}}$ becomes steeper for
larger line strengths
and shallower for smaller ones in this way reducing somewhat the effects
of the cutoff at lower t but introducing additional curvature at larger t.
As a result the power law exponent $\alpha(t)$  fitted according to
Eq.\,\ref{alpha} is optical depth dependent over the full range of t.

For our calculation of stellar winds in a wide range of stellar parameters
and metallicities we have, in a first step, calculated an extensive grid
of line force multpliers ${\it M}$ as function of t and $\hat{n}$ at
pre-specified and fixed values of effective temperature T$_{eff}$. The
calculations, which are very similar to those carried out by
\cite{kudet98} and \cite{pul00}
use a program code developed by U. Springmann during his thesis work
(see \citealt{spr97}). The line data base used has been build up by
A. Pauldrach and M. Lennon during the past 15 years and is described in
\cite{pau98}. It consists of wavelengths, gf-values,
photoionization cross sections and collision strengths for a total of 149
ionization stages and 2.5 million lines (for a critical discussion of
completeness, see also \citealt{pul00}). Non-LTE occupation numbers are
calculated in an approximate way assuming for the ionization equilibrium
of ground-state occupation numbers

\begin{equation}
\fre{n_{i,j+1}n_{e}}{\fri{n_{i,j}}} =
W (\fre{T_{e}}{\fri{T_{rad}}})^{0.5}
\lbrack \fre{n_{i,j+1}n_{e}}{\fri{n_{i,j}}}\rbrack^{LTE}_{T_{rad}}
\lbrace \zeta + \eta + W(1 - \zeta - \eta) \rbrace,
\label{ionis}
\end{equation}

where $\zeta$ and $\eta$ are the fractions of recombination
processes leading directly to the ground state and metastable
levels, respectively. $T_{e}$ is the electron temperature in the
stellar wind (adopted to be 0.8 T$_{eff}$), $T_{rad}$ the
radiation temperature at the ionization frequency calculated from
line blanketed unified NLTE model atmospheres with spherical
extension and stellar winds (see \citealt{pau98}, \citealt{pau01}) and W
the geometrical dilution factor of the radiation field (adopted to
be 1/3 at the base of the wind around the critical point). The
term denoted with $LTE$ represents the Saha-formula with $T_{rad}$
chosen for the temperature. A detailed justification for the use
of Eq.\,\ref{ionis} is given by \cite{spr97} and \cite{pul00}. A
similar equation but without $\eta$-terms has been used by
\cite{schmu91} and \cite{schae94}.

The excitatition of metastable states relative to the groundstate is adopted
to be the equilibrium population with regard to $T_{rad}$

\begin{equation}
\fre{n_{u}}{\fri{n_{1}}} = \lbrack \fre{n_{u}}{\fri{n_{1}}} \rbrack
^{LTE}_{T_{rad}},
\label{excimeta}
\end{equation}

wheras all other levels excited directly from the ground-state or a
metastable level are assumed to have a diluted population

\begin{equation}
\fre{n_{u}}{\fri{n_{l}}} = W \lbrack \fre{n_{u}}{\fri{n_{l}}} \rbrack
^{LTE}_{T_{rad}}.
\label{exciallo}
\end{equation}

For discussion, see again \cite{spr97} and \cite{pul00}.

\begin{figure}
\begin{center}
\figurenum{1}
\epsscale{0.45}
\plotone{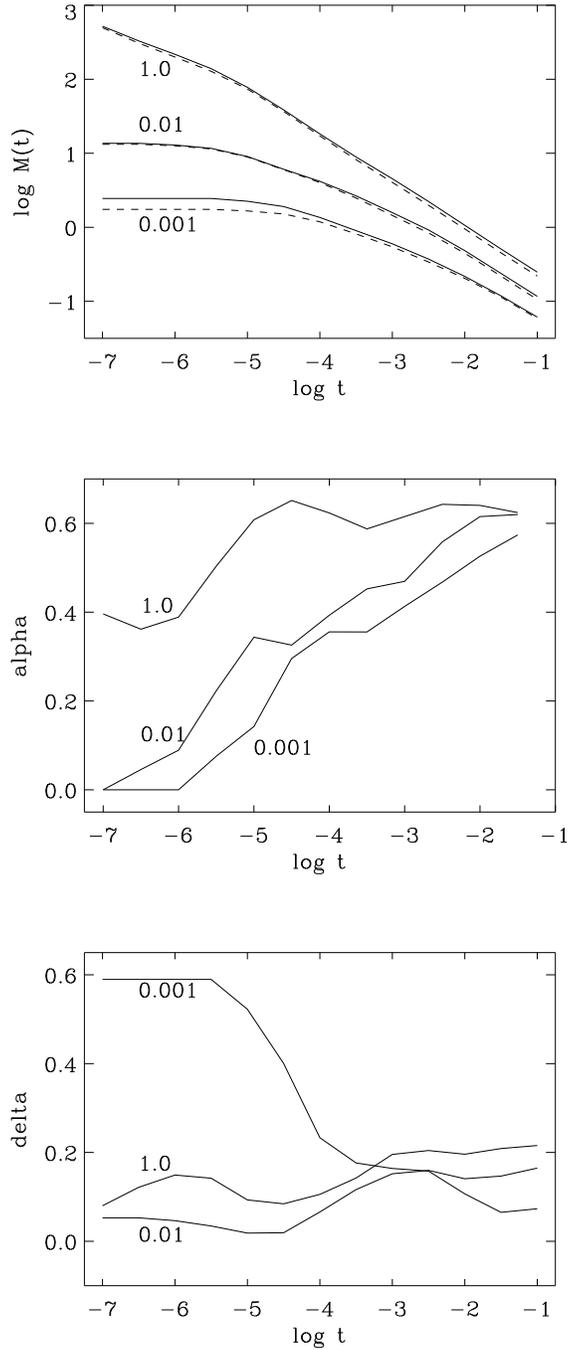}
\caption{Top: Line force multiplier ${\it M(t)}$ as a function of
optical depth parameter t for T$_{eff}$ = 50000K and metallicities
Z = 1.0, 0.01 and 0.001
Z$_{\odot}$. The solid curves correspond to a density log n$_{e}/W$ = 10.0,
for the dashed curves the density is 0.25 dex smaller.
Middle: Logarithmic derivative $\alpha$ of the line force
multipliers in the top panel as defined in Eq.\,\ref{alpha}.
Bottom: Logarithmic derivative $\delta$ of the line force
multipliers in Fig.\,\ref{fig1} as defined in Eq.\,\ref{delta}.
Note that both line force parameters, $\alpha$ and $\delta$, vary strongly as
function of optical depth for metallicites smaller than solar.
\label{fig1}}
\end{center}
\end{figure}

Fig.\,\ref{fig1} shows line force multipliers as a function of the optical
depth parameter calculated for different metal abundances. The deviation from
a simple power law at low optical depth parameter and the onset
of saturation of the line force multiplier can be easily recognized.
Consequently, the power law exponent $\alpha$ as defined in Eq.\,\ref{alpha}
depends on log t and becomes very small close to saturation. The influence
of the density parameter $\hat{n}$ is also
indicated in Fig.\,\ref{fig1}. If the parametrization of Eqs.\,\ref{lfm1}
were correct, then all dashed curves would have to be strictly
parallel to their solid counterparts in the double logarithmic plots. This
is obviously not the case. We conclude that the local fmp $\delta$ defined as

\begin{equation}
\delta(\hat{n},t) = -\fre{\partial log({\it M}(t,\hat{n}))}
{\fri{\partial log(\hat{n})}}
\label{delta}
\end{equation}

depends on density as well as on optical depth. The reason is that a
change of ionization does not only affect the normalisation of the
line-strength distribution function (Eq.\,\ref{lnorm}) but also its
slope (Eq.\,\ref{nk}). Fig.\,\ref{fig1} shows the optical depth parameter
dependence of $\delta$ calculated according to Eq.\,\ref{delta}. The high
values of $\delta$ for the low metallicity calculation at low log t are caused
by the fact that here the contributions from hydrogen and ionized helium are
already significant. \cite{pul00} have shown analytically that in such
a case $\delta$ can reach values close to unity.

\begin{figure}
\begin{center}
\figurenum{2}
\epsscale{1.0}
\plotone{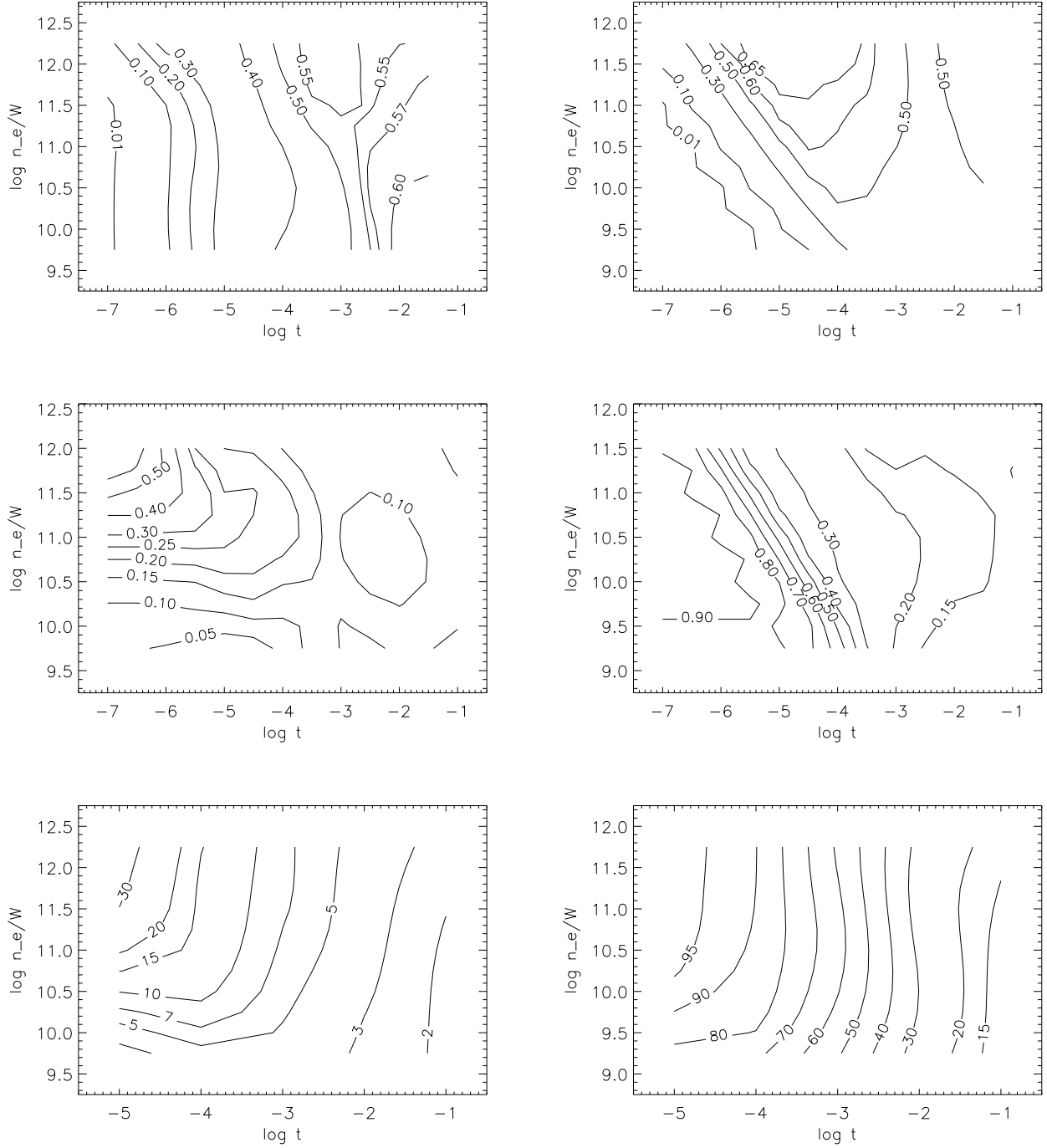}
\caption{Isocontours of line force parameters $\alpha$ (top panel)
and $\delta$ (middle panel) in the (log n$_{e}/W$, log t)-plane. The bottom
panel shows the isocontours of the HeII contribution to the line force (in
percent). All calculations are for T$_{eff}$ = 50000K.
Left panels: Z = 10$^{-2}$ Z$_{\odot}$;
right panels: Z = 10$^{-4}$ Z$_{\odot}$.
For discussion see text.
\label{fig2}}
\end{center}
\end{figure}

As an example for the full parameter dependence of the conventional
fmps, iso-contour diagrams of $\alpha$ and $\delta$
in the (log t, log n$_{e}$/W)-plane are given in  Fig.\,\ref{fig2}. Obviously
$\alpha$ is a strong function of optical depth at all densities and
approaches zero at low values of log t. The parameter $\delta$ depends
on both, optical
depth and density and comes close to unity for small optical depths in the
case of extremely low metallicity, where the contribution of the
HeII lines becomes sigificant.

Stellar wind models as the ones to be calculated in section 6 cover the whole
(log t, log n$_{e}$/W)-plane as displayed in the Fig.\,\ref{fig2}. Typically,
an individual model follows a diagonal trajectory through the
(log t, log n$_{e}$/W)-plane starting somewhere at the upper right and ending
with log n$_{e}$/W and log t smaller by 1.5 and 2.5, respectively. Depending
on the stellar parameters and the resulting mass-loss rate the trajectories
of the individual models are shifted relative to each other. In consequence,
the use of constant force multipliers to describe the radiative line force
can become quite inaccurate for a complete model set and even within one
individual model.

\begin{figure}
\begin{center}
\figurenum{3}
\epsscale{0.9}
\plotone{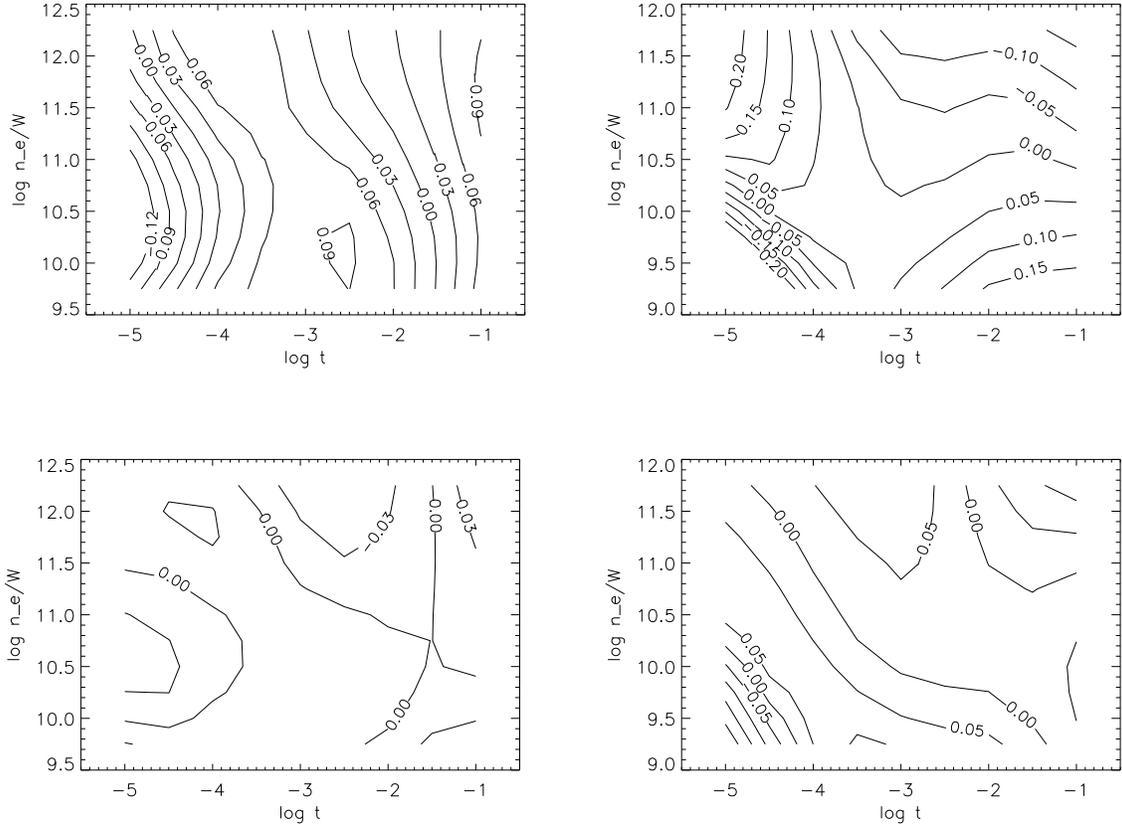}
\caption{Top: Isocontours of the difference between
log ${\it M(t)}$
and log ${\it M(t)}_{fit}$, where the latter is calculated with constant
values of $\alpha$ and $\delta$ according to Eq.\,\ref{lfm1}.
Bottom: The same as top, but now with
log ${\it M(t)}_{fit}$ calculated according to Eq.\,\ref{fmp3}.
Left: Z = 10$^{-2}$ Z$_{\odot}$; right: Z = 10$^{-4}$ Z$_{\odot}$.
\label{fig3}}
\end{center}
\end{figure}

As expected from Fig.\,\ref{fig1} and \,\ref{fig2}, Eq.\,\ref{lfm1} fails
badly to reproduce the line force
multiplier ${\it M}$ in the full parameter plane. This is demonstrated
by Fig.\,\ref{fig3}, where constant average values for $\alpha$ and
$\delta$ obtained from
multiple regression fits over the entire (log t, log n$_{e}$/W)-plane are used.

The fact that the force multiplier parameters depend on density and optical
depth parameter makes the simple calculations of line driven wind structures
as suggested by \cite{kud89} more difficult than previously
thought. Only in cases where the variations of  $\alpha$ and $\delta$ are
small or where it is sufficient to work with average values is it possible
to apply this concept. Otherwise, one has to deal with the variability of these
numbers as the elaborate and time consuming stellar wind codes do
\cite{pau94}.

Following \cite{kudet98} we have worked
out a solution to the problem which still allows a quick computation of
a large number of stellar wind models.
The simplest higher order approach of a fit formula for the force multiplier
is to assume that both $\alpha$ and $\delta$ depend linearly on $\log t$ and
$\log n_{e}/W$. With this assumption one obtains a new parametrization of
the form

\begin{equation}
\log {\it M}(t) = \log \hat{k} -\alpha_{o}(1+\alpha_{1}\log t)\log t +
\delta_{o}(1+\delta_{1}\log \hat{n})\log \hat{n} + \gamma \log t \log \hat{n}
\label{fmp3}
\end{equation}

This new parametrization gives a much more accurate representation of the line
force multiplier ${\it M}(t)$ at every effective temperature over the full
range in log t and log n$_{e}$/W. Fig.\,\ref{fig3} shows an example how the
accuracy of the representation of ${\it M}(t)$ is improved by
Eq.\,\ref{fmp3}. The new force multiplier parameters are compiled in
Table\,\ref{fmpsmet}.

The new parametrization, however, is only the first step in the solution to the
problem. The next and more difficult one is to achieve hydrodynamic stellar
wind solutions with the new representation of the radiative line force
similar to the one developed by \cite{kud89} but allowing for
depth dependent force multiplier parameters of the above form. This step is
undertaken in the following section.

\section{The equations of line driven winds with depth dependent force
multiplier parameters}

\subsection{The equation of motion}

In this section we develop a fast algorithm to calculate stellar wind
structures and mass-loss rates from the equation of motion (Eq.\,\ref{eom1})
using a radiative line acceleration parametrized in the form of
Eq.\,\ref{fmp3}. Our starting point is a formulation analogous to
\cite{kud89}, who restricted themselves to the case of an
isothermal wind which is a good approximation as far as the dynamics are
concerned (see \citealt{pau86}). Then the gas pressure is given
by

\begin{equation}
P_{gas}(r) = v^{2}_{s} \rho(r),
\label{pgas}
\end{equation}

where $v_{s}$, the isothermal sound speed, is constant throughout the wind.
We introduce as the geometrical depth variable

\begin{equation}
u = R_{\ast}/r
\label{u}
\end{equation}

(note that \citealt{kud89} have used x=1/u as depth variable; this is
the only difference with regard to the conceptual formulation of equations
and the definition of variables; for all other quantities not explicitely
defined refer to their paper) and as a quantity describing the velocity
gradient

\begin{equation}
y = -\fre{R_{\ast}}{\fri{2}} \fre{dv^{2}}{\fri{du}} .
\label{y}
\end{equation}

Then, the finite cone angle correction factor is given by

\begin{equation}
CF(u,y,v^{2}) = \fre{1}{\fri{1+\alpha(y)}} \fre{1}{\fri{\lambda}}
\lbrace 1 - (1-\lambda)^{1+\alpha} \rbrace,
\label{CF}
\end{equation}

where

\begin{equation}
\lambda(u,y,v^{2}) = u^{2} (1 - h), \qquad \qquad
h(u,y,v^{2}) = R_{\ast} \fre{v^{2}}{\fri{u}} \fre{1}{\fri{y}}.
\label{lambda}
\end{equation}

$R_{\ast}$ is the photospheric radius taken at a prespecified optical depth
in the visual continuum (see below). With these definitions and
Eqs.\,\ref{econt} and \ref{eom1} we obtain the non-linear implicit
differential equation for the velocity field

\begin{equation}
F(u,y,v^{2}) = C(y)f(u,y,v^{2})y^{\alpha(y)} - Aa - yb \equiv 0
\label{eom2}
\end{equation}

with

\begin{equation}
a = 1 - \fre{v_{s}^{2}}{\fri{v_{esc}^{2}}}\fre{4}{\fri{u}}, \qquad \qquad
b = 1 - \fre{v_{s}^{2}}{\fri{v^{2}}}.
\label{a}
\end{equation}

This equation of motion has the same structure as Eq. (5) in
\cite{kud89}, except that C and $\alpha$ are now variable and described by

\begin{equation}
\alpha = \alpha_{0} + \alpha_{0}\alpha_{1}(log(C_{t}/y)), \qquad \qquad
C_{t} = \fre{s_{e}v_{th}}{\fri{4\pi}} \dot{M}.
\label{alphay}
\end{equation}

and

\begin{equation}
C(y) = C_{0} C_{t}^{-\alpha(y)}, \qquad \qquad
C_{0} = \fre{s_{e}L}{\fri{4c\pi}} \hat{k}.
\label{C}
\end{equation}

with s$_{e}$ the electron scattering absorption coefficient
divided by the mass density $\rho$, $v_{th}$ the thermal velocity
of the protons and $\dot{M}$ the rate of mass-loss.

The function f is a product of CF and the function g

\begin{equation}
f(u,y,v^{2}) = CF(u,y,v^{2}) g(u,v^{2}), \qquad \qquad
g(u,v^{2}) = \hat{n}^{\delta(y,\hat{n})},
\label{f}
\end{equation}

where $\hat{n}$ as function of depth is calculated by

\begin{equation}
\hat{n}(u,v^{2}) = C_{\dot{M}} \fre{u^{2}}{\fri{W(u)}} \fre{1}{\fri{v}},
\label{nn}
\end{equation}

and

\begin{equation}
W(u) = \fre{1}{\fri{2}}(1-(1-u^{2})^{0.5}), \qquad \qquad
C_{\dot{M}}  = \dot{M} \fre{s_{e}}{\fri{R_{\ast}^{2}4\pi}} \cdot 1.503
\cdot 10^{24}.
\label{W}
\end{equation}

The force multiplier $\delta$ is also depth dependent through

\begin{equation}
\delta(y,\hat{n})  = \delta_{0} + \delta_{0}\delta_{1}log(\hat{n}) +
\gamma(log(C_{t}/y)),
\label{deltan}
\end{equation}

Finally, the constant A is

\begin{equation}
A = v^{2}_{esc} \fre{R_{\ast}}{\fri{2}}, \qquad \qquad
v_{esc} = (2g_{\ast}(1-\Gamma) R_{\ast})^{0.5}.
\label{A}
\end{equation}

$v_{esc} $ is the escape velocity from the stellar photosphere and
$g_{\ast}$ the photospheric gravity.

\subsection{Solution of the equation of motion}

As mentioned already in the previous subsection, the structure of
Eq.\,\ref{eom2} is identical to Eq. (5) in \cite{kud89},
except that C, $\alpha$ and $\delta$ are now variable. At a given
depth point in the stellar wind characterized by the radial co-ordinate
$u_{o}$ and the velocity $v(u_{o})$ Eq.\,\ref{eom2}
can be used as a non-linear algebraic equation to
calculate y. Fig. 2 of \cite{kud89} demonstrates that there
are usually two solutions which can be easily obtained numerically. Inward of
the critical point (see next subsection) the smaller of the two has to be
used, whereas outward the larger one is correct. At the critical point,
there is only one solution.

The solution y of Eq.\,\ref{eom2} at a given depth point in the wind
together with Eq.\,\ref{y} can then be used for an integration of the
equation of motion using

\begin{equation}
v^{2}(u) = v^{2}(u_{o}) + \fre{2}{\fri{R_{\ast}}} \int_{u}^{u_{o}}ydu
\label{eomint}
\end{equation}

The starting point for the integration will be the critical point.

\subsection{The singularity and regularity conditions at the critical point}

The function F of Eq.\,\ref{eom2} has a singularity at its critical point
$u_{c}$  (\citealt{cas75}; \citealt{pau86};
\citealt{kud89}). At $u_{c}$ exists only one unique value $C_{c}$ of
the function $C(\alpha)$, for which a smooth transition from very low
velocities in the photospheres to very high velocities of the order of or
larger than the photospheric escape velocities (as observed for O-stars) is
possible. $C_{c}$, the critical velocity $v_{c}$ and the critical velocity
gradient $y_{c}$ are determined from the singularity and the regularity
condition

\begin{equation}
\fre{\partial F}{\fri{\partial y}} = 0, \qquad \qquad
\fre{\partial F}{\fri{\partial u}} +
\fre{\partial F}{\fri{\partial v^{2}}} \fre{\partial v^{2}}{\fri{\partial u}}
= 0
\label{singcond1}
\end{equation}

together with the equation of motion at the critical point. After a lengthy
calculation one obtains from the first equation in Eq.\,\ref{singcond1}
the velocity gradient at the critical point

\begin{equation}
y_{c} = \fre{\alpha_{eff}^{c}}{\fri{1-\alpha_{eff}^{c}}}
p(u_{c},y_{c},v^{2}_{c}) A
\label{singcond2}
\end{equation}

with

\begin{equation}
p = \fre{a}{\fri{b}}, \qquad \qquad
\alpha_{eff} = \alpha D_{\alpha}(u,y,v,\alpha).
\label{p}
\end{equation}

The function $D_{\alpha}$, which contains the terms resulting from the depth
dependence of the force multipliers and the finite cone angle correction
factor with regard to y, is given in the Appendix. If the force multipliers
were constant (i.e. $\alpha_{1}, \delta_{1} = 0$) and the finite cone angle
correction factor equal to unity (the photon radial streaming approximation
by \citealt{cas75}), then $D_{\alpha} = 1$ would result. The function p is
of the order of unity and also given in the Appendix.

The second equation of Eq.\,\ref{singcond1} yields an expression for the
critical velocity through

\begin{equation}
v^{2}_{c} = v_{s}^{2} + \Delta v^{2}_{reg} (u_{c}, y_{c},
v^{2}_{c}, C_{c}). \label{regcond2}
\end{equation}

The function $\Delta v^{2}_{reg}$ is also provided in the
Appendix. From Eq.\,\ref{eom2} and \ref{singcond1} we derive

\begin{equation}
C_{c} = y_{c}^{-\alpha_{c}} \fre{A}{\fri{f_{c}}} \fre{1}{\fri{1-\alpha_{eff}}}.
\label{Ccrit}
\end{equation}

For a given value of the coordinate $u_{c}$ of the critical point the system
of Eqs.\,\ref{singcond2}, \ref{regcond2} and \ref{Ccrit} can be solved by
iteration (see below) yielding initial conditions for the integration of
the equation of motion (Eq.\,\ref{eomint}) inward and outward from the
critical point.

Once $C_{c}$ is determined, it can be used to determine the mass-loss rate
$\dot{M}$ as the eigenvalue of the problem by combining
Eqs.\,\ref{C}, \ref{W} and \ref{Ccrit}

\begin{equation}
\dot{M} = \lbrace \fre{C_{\dot{M}}}{\fri{\dot{M}}}
\fre{u_{c}^{2}}{\fri{v_{c}}}
\fre{10^{-11}}{\fri{W(u_{c})}}\rbrace
^{\delta_{c}/ \alpha_{c}^{\prime}} \cdot
CF(u_{c},y_{c},v^{2}_{c})^{1 / \alpha_{c}^{\prime}} \cdot
\dot{M}_{CAK}^{\alpha_{c} / \alpha_{c}^{\prime}}, \qquad \qquad
\alpha^{\prime} = \alpha - \delta,
\label{mdot}
\end{equation}

\begin{equation}
\dot{M}_{CAK} = C_{0}^{1 / \alpha_{c}} \cdot
\fre{\dot{M}}{\fri{C_{t}}} \cdot
\fre{\alpha_{eff}^{c}}{\fri{b_{c}}} \cdot
\lbrace \fre{1-\alpha_{eff}^{c}}{\fri{Aa_{c}}}
\rbrace ^{\alpha_{c} / (1-\alpha_{c})}  .
\label{mcak}
\end{equation}

The structure of Eq.\,\ref{mdot} is identical to Eq. 65 of \cite{kud89},
although it contains different terms resulting from the
force multiplier depth dependence and the exact treatment of the function f.
The scaling relations of the mass-loss rate with regard to luminosity L,
stellar mass and distance to the Eddington limit remain the same,
qualitatively, although in practice the implicit dependence of $\alpha, \delta$
and $\alpha_{eff}$ on these quantities may induce quantitative changes.
Note that $C_{\dot{M}}/\dot{M}$ and $\dot{M}/C_{t}$ are independent of
$\dot{M}$.
The iterative determination of y$_{c}$, $v_{c}$, $C_{c}$ and $\dot{M}$ from
Eqs.\,\ref{eom2}, \ref{singcond2}, \ref{regcond2}, \ref{Ccrit} and
\ref{mdot} is made for a given value of $u_{c}$, the coordinate of the
critical point. The calculation of $u_{c}$ itself is described in the
next subsection.

\subsection{The location of the critical point}

Following \cite{cas75} and \cite{pau86} we calculate the
location of the critical point from the condition that the photospheric
radius $R_{\ast}$ must correspond to a pre-specified monochromatic optical
depth $\tau_{\lambda}^{Phot}$

\begin{equation}
\tau_{\lambda}^{Phot} = \int^{\infty}_{R_{\ast}} \kappa_{\lambda} dr,
\label{tau1}
\end{equation}

where the wavelength $\lambda$ is taken to be 5500 A,
corresponding to the V-band photometry. A reasonable value for
$\tau_{\lambda}^{Phot}$ is 2/3. $\kappa_{\lambda}$ is the
monochromatic absorption coefficient, which can be written as

\begin{equation}
\kappa_{\lambda} = s_{e} \rho (1+a_{\lambda}\rho).
\label{kappa}
\end{equation}

The second term of Eq.\,\ref{kappa} contains the contributions of bound-free
and free-free absorption in addition to electron scattering as given by
the first term. Assuming LTE and allowing for the contributions of
hydrogen and helium only (a good approximation for the V-band
continuous opacity of hot stars), $a_{\lambda}$ is a function of electron
temperature and helium abundance and can be easily calculated. (Note that older
work by \citealt{cas75}, \citealt{pau86} and \citealt{kud89} neglects the
contribution of bound-free and free-free opacities).

Using Eqs.\,\ref{econt} and \ref{u} $\tau_{\lambda}^{Phot}$ can be
expressed as

\begin{equation}
\tau_{\lambda}^{Phot} = s_{e} \fre{\dot{M}}{\fri{4\pi}}
\fre{1}{\fri{R_{\ast}}} I(\dot{M},R_{\ast}).
\label{tau2}
\end{equation}

The function $I(\dot{M},R_{\ast})$ contains two integrals over the velocity
field $v(u)$

\begin{equation}
I = \int^{1}_{0}\fre{du}{\fri{v}} + \fre{\dot{M}}{\fri{4\pi}}
\fre{a_{\lambda}}{\fri{R_{\ast}^{2}}} \int^{1}_{0}u^{2}\fre{du}{\fri{v^{2}}}.
\label{I}
\end{equation}

For given $\dot{M}$, $y_{c}$ and $v_{c}$ the value of I depends strongly on
$u_{c}$ as it defines the density $\rho_{c}$ at the critical point via the
equation of continuity and the transition from a wind into a hydrostatic
stratification in deeper layers. Eq.\,\ref{tau2} can, therefore, be used to
iterate for the correct value of $u_{c}$ to match the pre-specified value
of $\tau_{\lambda}^{Phot}$.

\subsection{The full iteration cycle}

Because of the mutual dependence on u,y,$v$ and $\dot{M}$ of many of the
functions introduced in the previous subsections a careful and complex
iteration procedure is needed to solve the system of equations. The following
scheme proved to be stable with good convergence over a wide range of
stellar parameters.

We use the algorithm by \cite{kud89} for $\alpha=0.55$ and
$\delta=0.1$ to obtain starting values for $u_{c}$, $y_{c}$, $v_{c}$,
$C_{c}$ and $\dot{M}$. For $u_{c}$ fixed we then apply three
interlocking iteration cycles.
The first one is the innermost iteration cycle and uses Eq.\,\ref{regcond2}
to iterate for $v_{c}$ with $y_{c}$, $C_{c}$, $\alpha_{c}$ and $\delta_{c}$
fixed. Once convergence for $v_{c}$ in the innermost cycle is achieved, we
start the next cycle by calculating $y_{c}$ from
Eq.\,\ref{singcond2} with $v_{c}$, $C_{c}$, $\alpha_{c}$ and
$\delta_{c}$ fixed. For every new value of $y_{c}$ in this second
cycle we use again cycle one for $v_{c}$, until both $y_{c}$ and $v_{c}$
converge. Then we start the outermost cycle three, which calculates new
values of $C_{c}$, $\dot{M}$, $\alpha_{c}$ and $\delta_{c}$ and leads to a
new value of $y_{c}$. With this new value of $y_{c}$ the two inner iteration
cycles one and two are started again and the whole procedure is iterated,
until full convergence of $y_{c}$, $v_{c}$, $C_{c}$, $\dot{M}$, $\alpha_{c}$
and $\delta_{c}$ is obtained. Then we solve the equation of motion in both
directions from the critical point to obtain the full velocity field $v(u)$.
Integration of the velocity field (Eq.\,\ref{tau2}) yields a photospheric
optical depth and the comparison with the pre-specified value leads to a new
estimate for $u_{c}$. With this new value the three inner iteration cycles
are started again and the procedure is repeated, until the correct value
for $u_{c}$ is found.

Although this iteration procedure looks very complicated and time consuming,
it is straightforward to implement and takes only a few seconds on a
workstation to converge.

\section{A test of the wind models. O-stars in the Galaxy and the SMC}

The new approach to calculate wind models developed in the foregoing sections
needs to be tested observationally before it can be applied to predict
wind properties of very massive stars at low metallicity. The ideal objects
for this purpose are the most luminous and most massive O-stars in the
Galaxy and the Magellanic Clouds, the latter because of the reduced
metallicity in these galaxies.

As has been demonstrated by \cite{pul96} and \cite{kud89}, the
best way to discuss the strengths of winds of hot stars is in terms
of the wind momentum - luminosity relationship (WLR). The theory of
radiation driven winds predicts that for O-stars the
``modified stellar wind momentum''

\begin{equation}
D_{\rm mom} = \dot{M} v_{\infty} (R_{\ast}/R_{\odot})^{0.5}
\label{modmom}
\end{equation}

depends mostly on stellar luminosity and much less on other stellar parameters.
This prediction has been confirmed very convincingly by the empirical
spectroscopic diagnostics of O-star winds. Fig. 4 shows the observed modified
wind momenta of galactic O-stars
and Central Stars of Planetary Nebulae (CSPN) as taken from \cite{kud00}.
Both the O-supergiants and
O-giants and -dwarfs follow rather tight relationships, which when
extrapolated towards lower luminosities coincide with the observed wind
momenta of CSPN (for the two dwarfs falling off the relationship, see
discussion in \citealt{pul96}).

Fig. 4 also shows the results of the model calculations using our new approach
with the force multipliers given in Table 1. The models for O-stars
have effective temperatures of 50000 and 40000 K, respectively. For the
supergiants we have adopted gravities log g between 3.75 to 3.95 at 50000K
and 3.35 to 3.50 at 40000K. Giants and dwarfs have larger gravities, 4.10
to 4.20 at 50000K and 3.7 to 4.0, respectively, at 40000K. These values
coincide roughly with the stellar parameters observed. To calculate
the winds of the CSPN the core-mass luminosity relationship for post-AGB
has been adopted to estimate stellar masses and gravities at the
corresponding luminosities. Solar metallicity was used for all the wind
models of galactic objects.

The agreement with the regression curves resulting from the
observations is satisfying though not perfect. The models for the
supergiants produce slightly too weak winds and the opposite is
the case for the dwarfs and giants. However, the general trend is
reproduced very well. This indicates that our algorithm to
calculate radiative line forces produces fmps of the right order
of magnitude. In addition, the concept of variable fmps leads to a
more pronounced difference between wind models close to the
Eddington-limit and on the main sequence in better agreement with
the observations.

In addition, Fig. 4 compares wind momenta of O-stars in the Galaxy
and the metal poor SMC showing that winds are significantly weaker
at lower metallicities (again from \citealt{kud00}). The calculations
reproduce this trend, as is shown by Fig 4 as well. The stellar
parameters used for the SMC calculations are identical to those
for the Galaxy, except that Z = 0.2 Z$_{\odot}$ has been adopted
for the metallicity following the results obtained by \cite{has98}
from the analysis of HST spectra of O-stars in the SMC. We have
restricted the SMC calculations to dwarfs and giants because all
but the most luminous object in Fig. 10a belong to these
luminosity classes. As for the Milky Way there are a few (three)
objects at lower luminosities, which fall off the relationship and
the theory is not able to explain their wind momenta (but see
Kudritzki and Puls, 2000, for discussion). In general, because of
the small number of objects studied so far, the WLR in the SMC is
not as well defined as in the Milky Way. More spectroscopic work
is needed to improve the situation.

A comparison between calculated and observed terminal velocities of the stellar
winds is another important test of the theory. For galactic O-stars this is
carried out in Fig.\,\ref{fig5}, which displays terminal velocities as a
function of
photospheric escape velocities, since the theory predicts that both are
correlated to first order. The result of the test is quite encouraging,
although the observed terminal velocities are on the average somewhat
higher than the calculated ones by 5 to 10 percent. Whether this small
discrepancy reflects a deficiency of the theory or a systematic effect
resulting from the determination of the observed escape probabilities, is an
open question which is beyond the scope of this paper.

Metallicity does also affect the terminal velocities of stellar winds in a
systematic way so that winds become slower with decreasing metal abundance,
as investigations of O-stars in the Magellanic Clouds have revealed (refer
to \citealt{kud00}, for references and a compilation of results).
Fig.\,\ref{fig5} shows that this effect is also reproduced by the theory.

\section{Wind models for very massive stars at low metallicity}

After the new concept to calculate stellar wind structures with variable
force multipliers has been introduced and tested by comparing with the
observed wind properties of O-stars in the Galaxy and the SMC, we are now
ready for an application on very massive stars. The purpose of this first
study is to provide an estimate about the strengths of stellar winds
at very low metallicity for very massive hot stars in a mass range roughly
between 100 to 300 M$_{\odot}$. We concentrate on an
effective temperature range comparable to the hottest and most massive
observed O-stars, i.e. 60000K to 40000K, which is only a mild extrapolation
away from a stellar parameter regime, where the theory has been tested at
galactic and SMC metallicity. We are fully aware of the fact that at the low
metallicities used the zero age main sequences of stars in this mass range
are shifted to higher effective temperatures ($\approx$ 75000K, see
\citealt{bar01}) than accounted for in our
calculations. However, at this stage we do not aim at a comprehensive
description of stellar winds of metal poor very massive stars through all
stages of their evolution. Instead, we restrict ourselves to a temperature
regime, where the winds of massive stars at normal metallicity are well
understood. We defer a more complete description connected with realistic
evolutionary tracks to a second paper.

In a first step, following the discussion in section 2, we will
investigate the strengths of low metallicity winds as a function
from the distance to the Eddington limit.  We will then define a
simple schematic grid of stellar parameters, which will allow to
investigate the systematic behavior of mass-loss rates, wind
velocities, wind momenta and wind energies. We will use the full
set of wind models calculated for this grid to provide simple
analytical fit formulae for stellar wind properties as a function
of stellar parameters and metallicity.

\subsection{Low metallicity winds and the distance to the Eddington - Limit}

In Section 2 we concluded from simple analytical considerations that
line driven winds at very low metallicity can only be maintained if stars
are close enough to the Eddington-limit so that the effective gravity becomes
negative somewhere out in the wind. To investigate this effect by using
the full stellar wind equations and a realistic line acceleration we have
calculated sequences of models at constant luminosity and effective
temperature but with the different stellar masses and, therefore, different
distances to the Eddington-limit. The result is displayed in
Fig.\,\ref{fig6}. The small and smooth dependence of the modified stellar
wind momentum on $\Gamma$ for solar and SMC metallicity is well understood in
terms of the discussion given by \cite{pul96}. For constant luminosity
the scaling relations of line driven winds predict

\begin{equation}
log D_{\rm mom} \propto (3/2-1/\alpha^{\prime}) \cdot  log M(1-\Gamma)
 \propto (3/2-1/\alpha^{\prime}) \cdot log (1/\Gamma-1),
\label{momalph}
\end{equation}

which means that the stellar wind momenta
should decrease slightly with decreasing $\Gamma$, if
$\alpha^{\prime} \leq 2/3$. The average force multiplier parameters of O-stars
in the Galaxy and the Magellanic Clouds lead to values of $\alpha^{\prime}$
between 0.50 to 0.55, which explains the $\Gamma$-dependence of wind momenta
in the corresponding range of metallicities. For Z/Z$_{\odot}$ = 0.01 the slope
with $\Gamma$ becomes much stronger corresponding to an average
$\alpha^{\prime}$ of the order of 0.4. It is, however, still possible to find
wind solutions in the
full range of $\Gamma$ appropriate for the luminosity adopted (note that
$\Gamma$ = 0.4 corresponds to 119 M$_{\odot}$ at log L/L$_{\odot}$ = 6.26).
For smaller metallicities the slope becomes even steeper and winds become
very weak as soon as a certain threshold in $\Gamma$ is reached. For lower
$\Gamma$, no wind solutions could be found,
confirming qualitatively the analytical estimate of section 2.

\subsection{Adopted stellar parameters for very massive objects}

The goal of this paper is to investigate systematically the role of winds as
function of metallicity and stellar parameters for hot stars in a mass range
between 100 to 300 M$_{\odot}$. To do this consistently, i.e. by a combination
of evolutionary tracks with stellar wind models is complicated because of the
mutual dependence. For a given stellar mass, the evolution of a massive star,
i.e. its location in the HRD depends strongly on metallicity but also on the
strength of mass-loss. On the other hand, the properties of stellar winds
depend also very strongly on the stellar parameters adopted and, of course,
on metallicity. To disentangle this mutual dependence we proceed in a
straightforward way. We define stellar parameters independent of metallicity
for very simplified evolutionary sequences. In this way, we ignore the
detailed effects of metallicity on the stellar evolution but we will be able
to discuss its direct influence on stellar winds. Using the full information
of wind models calculated for the grid of different stellar parameters and the
different sets of metallicities we will then be able to provide fit formulae
which can be used in conjunction with consistent evolutionary calculations in
the future.

Our starting point is the paper by \cite{schal92} which provides
stellar models from 0.8 to 120 M$_{\odot}$ at solar and 1/20 solar metallicity.
For our considerations we concentrate on their low metallicity models, which
at their high mass end lead to a zero age main sequence mass-luminosity
relationship of

\begin{equation}
log (L/L_{\odot})^{ZAMS} = 6.095 +1.53 \cdot \lbrace log (M/M_{\odot}) - 2
\rbrace
\label{zams}
\end{equation}

Evolving away from the ZAMS these objects gain roughly 0.2 dex in luminosity.
Thus, we adopt for the more advanced evolutionary state in the temperature
range of 60000 to 40000 K considered in this study

\begin{equation}
log (L/L_{\odot}) = log (L/L_{\odot})^{ZAMS} +0.2
\label{tams}
\end{equation}

Eq.\,\ref{zams} and \ref{tams} can then be used to calculate radii and wind
models at 40000, 50000 and 60000K effective temperature for the masses as
indicated in Fig.\,\ref{fig7}. A comparison with models of very massive stars
published in the literature (see, for instance, \citealt{bro01b},
\citealt{bar01}) shows that this simple extrapolation is quite reliable.

\subsection{Stellar wind properties}

Table\,\ref{windprop} gives a complete overview of the stellar wind properties
for every model calculated and provides terminal velocity, mass-loss rate and
modified wind momentum together with stellar parameters and metallicity. In the
following we discuss the most important systematic trends using the models at
T$_{eff}$ = 50000K as an example.

Fig.\,\ref{fig8} displays the modified stellar wind momentum as function of
luminosity for models at different metallicity. The simplified theory of
radiation driven winds with depth independent and luminosity independent
force multiplier parameters predicts a simple relationship of the form

\begin{equation}
log D_{\rm mom} \propto 1/\alpha^{\prime} \cdot  log L +
(1-\alpha^{\prime})/\alpha^{\prime} \cdot (Z/Z_{\odot}),
\label{momalz}
\end{equation}

(\citealt{kud89}, \citealt{pul96}, \citealt{kud00}).
Fig.\,\ref{fig8} confirms that, in principle, such a relationship continues to
exist for the more elaborated force multiplier approach developed in this
paper, however the effective value of $\alpha^{\prime}$ depends now on both,
luminosity and metallicity. In particular, it decreases very significantly
with metallicity (see also discussion in section 6.1) so that the dependence
of wind momentum on luminosity becomes much steeper with decreasing
metallicity. This trend becomes also obvious in Fig.\,\ref{fig9}, where the
stellar wind properties are plotted as function of metallicity. Mass-loss,
modified wind momentum and wind energy decrease stronger than a simple power
law with metallicities beyond $Z/Z_{\odot}$ = 0.01.

In the same framework of simplified wind theory with constant
force multipliers the terminal velocity of the stellar wind is
related to the escape velocity from the stellar surface via the
proportionality (\citealt{kud00})

\begin{equation}
v_{\infty}/v_{esc} \propto \alpha/(1-\alpha) \cdot exp(-2\delta) .
\label{vinfty}
\end{equation}

From Fig.\,\ref{fig9} we have learned that the effective value of
$\alpha^{\prime}$ and, thus, also $\alpha$ is decreasing at low metallicities.
According to Eq.\,\ref{vinfty} we, therefore, expect smaller ratios of
$v_{\infty}/v_{esc}$ at the low metallicity end as the result of our
calculations, which is confirmed by Fig.\,\ref{fig9}. In addition, we find
that the ratio does also depend on luminosity and becomes smaller with
decreasing luminosity.

\subsection{An analytical fit of mass-loss rates to stellar parameters and
metallicity}

Stellar evolution calculations for very massive stars need to include the
effects of mass-loss as soon as the mass-loss rates are high enough to reduce
the total stellar mass significantly during the different phases of stellar
evolution. While in the range of solar and Magellanic Cloud metallicities
the standard formulae resulting from the theory of line driven winds or fits
to the observed data are usually applied, nothing comparable is available
for the stellar parameter and metallicity domain investigated here. We have,
therefore, developed a simple analytical fit-formula to our numerical results,
which provides mass-loss rates of line driven winds. It can easily be combined
with stellar evolution calculations or used to estimate the energy and momentum
input of very massive low metallicity stars to the interstellar medium.

\begin{equation}
log \dot{M} = Q([Z]) = q_{1} ([Z]-[Z]_{min})^{0.5} + Q_{min},
\qquad \qquad q_{1} = (Q_{0}-Q_{min})(-[Z]_{min})^{-0.5}.
\label{mdz}
\end{equation}

[Z] is defined as

\begin{equation}
[Z] = log (Z/Z_{\odot})
\label{Z}
\end{equation}

$[Z]_{min}, Q_{min}, Q_{0}$ depend on the luminosity through a simple
polynomial formula

\begin{equation}
y = a_{0} +  a_{1}\tilde{L} + a_{2}\tilde{L}^{2},
\qquad \qquad \tilde{L} = log (L/L_{\odot}) - 6.0.
\label{mdzcoeff}
\end{equation}

The coefficients are $a_{0}, a_{1}, a_{2}$ for the fits of
y=$[Z]_{min}, Q_{min}, Q_{0}$ are given in Table \,\ref{mdtab}.

\subsection{Decoupling of radiatively accelerated ions at low metallicity}

The key process of line driven winds is the transfer of radiative
momentum absorbed in spectral line transitions of metal ions to
the bulk mass of ionized hydrogen and helium, which because of the
lack of enough strong line transitions is not much driven directly
by radiation. For the strong and relatively dense winds of O-stars
at solar metallicity the transfer mechanism is provided by Coulomb
collisions, which keep the metal and the hydrogen/helium ions
tightly coupled together, as has been shown by \cite{cas76}.
However, for weak winds with low mass-loss rates and
correspondingly low densities the lower collision rates can lead
to a decoupling of the ions from the bulk plasma and produce a
``ion runaway'' reducing mass-loss rate and wind momentum
significantly (\citealt{spr92}, \citealt{bab95}, \citealt{por95},
and \citealt{por99}). The effect of a runaway
was put into question recently by \cite{krt00}, \cite{krt01},
who were the first to derive complete quantitative solutions for
two-component (ions and passive plasma) steady state line driven
winds. They found that in the limit of low density winds both the
ions and the passive plasma adopt a solution of lower acceleration
and avoid the runaway. This very interesting result was then
challenged very recently by \cite{owo02}, who carried
out a time-dependent, linearized stability analysis of the
two-component solutions and found that a runaway is very likely to
happen in the wind flow, before the wind adopts the steady,
slow-acceleration solutions. Thus, the situation in the low
density limit of weak winds remains unclear at this point but
there is the clear potential that normal single-fluid solutions
might become unrealistic in this limit.

At the extremely low metallicities considered in this investigation the
single-fluid solutions might have crossed over into a parameter domain, where
their validity has become questionable, because the rate of Coulomb collisons
has dropped significantly because of extremely low mass-loss rates and the
very low abundance of ions. We have, therefore, used equation 16b
(corrected for a numerical mistake in their treatment of the Coulomb
Logarithm) of Springmann and Pauldrach and equation 22 of Owocki and Puls to
check, which of our models are in a critical parameter domain with regard to
a possible runaway.

For $Z/Z_{\odot}$ = 0.0001 the models with $log L/L_{\odot}$ = 6.91 might
suffer from a runaway, but only at velocities of the order of one half to one
third the terminal velocity and certainly larger than the critical velocity,
thus probably still little affected by two-component effects. The same is true
for some models with $Z/Z_{\odot}$ = 0.001 and $log L/L_{\odot} \le $ 6.42.
A more detailed investigation will be needed for these models, but we conclude
that in general the winds calculated in our model grid are only marginally
affected by ion decoupling.

\section{Ionizing fluxes and stellar spectra}

With the stellar wind structures and parameters specified in the
previous sections we can now calculate detailed atmospheric models
together with stellar energy distributions and synthetic spectra
for all the models in Table\,\ref{windprop} . For this purpose we
used the approach of "Unified Model Atmospheres" as developed at
Munich University Observatory over the last 15 years. These model
atmospheres are in NLTE, radiative equilibrium, spherically
extended and include the effects of stellar winds (see \citealt{gab89},
for the original introduction of the concept and
\citealt{pau94}, for a first version including metal lines).
The most recent step of this development by \cite{pau01}
accounts for more than 4 millions of metal lines in NLTE
originating from more than 150 metal ions. Detailed atomic models
with accurate atomic data are set up for every ion, for which the
equations of statistical equilibrium are solved consistently and
simultaneously with the radiative transfer in each line and
ionization transition in a highly iterative algorithm. Multi-line
absorption is included in the radiative transfer and in the
radiative equilibrium, which means that the effects of
line-blanketing and -blocking are fully taken into account. After
convergence spectra and energy distributions are calculated
including all the spectral lines. The code is thus ideally suited
to demonstrate the transition from solar to very low metallicity.
In addition, despite the complexity of the atomic models and the
radiative transfer algorithms the code is extremely fast and
produces a converged model on a laptop or PC in a few hours. It is
public available and can be downloaded from the Munich University
Observatory website (http://www.usm.uni-muenchen.de/people/adi/adi.html).
For details we refer the reader to the original publication.

Fig.\,\ref{fig10} gives an example of the effects of metallicity
on the EUV and FUV spectral energy distribution. We have selected
two models at $T_{eff}$ = 60000K at two different luminosities,
log $L/L_{\odot}$ = 6.57 and 6.91, respectively. Longward of 228A,
in the H, HeI, OII, NeII, CIII ionizing continuum, the influence
of metal line opacity is very similar at both luminosities.
Increased metallicity decreases the emergent flux within the metal
lines because of enhanced line blocking but increases the flux
emitted in the continuum windows with reduced line opacity because
of the back-warming effects of line-blanketing. The balance
between the different influence of blocking and blanketing will,
therefore, determine how the ionizing properties of these stars
are affected by metallicity (see below).

Shortward of 228A, in the HeII ionizing continuum, metallicity has
a dramatic influence on the size of the HeII absorption edge and,
thus, on the ionizing flux. However, this influence is less
related to the effects of line-blanketing and -blocking rather
than to the strengths of the stellar winds correlated with
metallicity (see \citealt{gab89}, \citealt{gab91}, \citealt{gab92} for a
detailed explanation).

Fig.\,\ref{fig11} summarizes the effects of metallicity on the
ionizing properties of very massive stars. For three effective
temperatures and two luminosities we show the number of emerging
ionizing photons per stellar surface element and unit time as a
function of metallicity. To characterize the wavelength dependence
of the ionizing radiation we display the number of photons being
able to ionize H (ionization edge at 911A), HeI (504A), OII
(353A), NeIII (303A), CIII (259A) and HeII (228A). The dependence
on metallicity of these photon numbers varies significantly with
the ionization edge approaching the limit of HeII ionization.
While the number of H photons remains almost constant and the
effects of metal line-blocking and -blanketing balance out over
the relatively wide spectral range from the hydrogen to the HeII
absorption edge, the number of CIII photons decreases strongly
with increasing metallicity, because line blocking dominates the
remaining wavelength interval towards the HeII absorption edge.

The HeII ionizing photons reflect a more complex behavior as
discussed above. At solar metallicity winds can become highly
optically thick in the HeII continuum and then the number of
ionizing photons drops dramatically. As soon as the winds become
weak enough, the velocity field induced ground-state de-population
(Gabler et al., 1989) sets in and increases the photon number
substantially. Then, with decreasing metallicity the winds become
weaker, which reduces the influence of the de-population effect.
Since the wind strengths depend strongly on stellar luminosity,
the detailed metallicity dependence of the HeII photons varies
with luminosity. Fig.\,\ref{fig12} gives examples for $T_{eff}$ =
60000K and 50000K, respectively. Thus, to predict the ionizing
flux in the HeII continuum appropriately for a very massive star
at a given metallicity  requires a calculation of the wind
parameters first.

The luminosity dependence of the ionizing photons for hydrogen and
neutral helium turns out to be very weak. To a very good
approximation the numbers displayed in Fig.\,\ref{fig11} are
representative. The luminosity effect for the OII, NeII photons is
somewhat larger but still small. The largest effects are found for
the CIII photons as displayed in Fig.\,\ref{fig12}.

It is also interesting to calculate synthetic FUV and UV spectra as a
function of metallicity. As we know well from IUE, HST, ORFEUS and
FUSE observations of massive stars in the Galaxy and the Large and
Small Magellanic Cloud, the observed spectra in this spectral
range are heavily blended by a dense forest of slightly wind
affected pseudo-photospheric metal absorption lines superimposed
by broad P Cygni and emission line profiles of strong lines formed
in the entire wind. Model atmosphere synthetic spectra are able to
reproduce the spectra nicely (\citealt{has98}, \citealt{dek98},
\citealt{ful00}, \citealt{pau01}), demonstrating
the reliability of the model atmosphere approach. The interesting
question to investigate is to find the metallicity range, where
the UV stellar wind features and the photospheric metal lines
absorption lines start to disappear. Fig.\,\ref{fig13} gives an
overview for log L/L$_{\odot}$ = 6.91, where the winds are relatively strong.
The stellar wind lines (NV $\lambda$ 1240, OV $\lambda$ 1371,
CIV $\lambda$ 1550, HeII $\lambda$ 1640) remain clearly
visible down to Z/Z$_{\odot}$ = 10$^{-2}$, where the absorption forest
has already started to disappear. For lower metallicity, the stellar wind
character of these lines disappears, but they are still detectable as
photospheric absorption or emission lines. This means that for starbursting
galaxies at very high redshift and possibly very low metallicities, as
eventually
observable with NGST in the IR, there is still diagnostic information
available to estimate chemical abundances and further properties such as
the Initial Mass Function and the star formation rate of the integrated stellar
population.

\section{Discussion and future work}

With our new approach to describe line driven stellar winds at
extremely low metallicity we were able to make first predictions
of stellar wind properties, ionizing fluxes and synthetic spectra
of a possible population of very massive stars in this range of
metallicity $Z/Z_{\odot}$. We have demonstrated that the normal
scaling laws, which predict stellar-mass loss rates and wind
momenta to decrease as a power law with $Z/Z_{\odot}$ break down
at a certain threshold and we have replaced the power-law by a
different fit-formula. We were able to disentangle the effects of
line-blocking and line-blanketing on the ionizing fluxes and found
that while the number of photons able to ionize hydrogen and
neutral helium is barely affected by metallicity (and stellar
luminosity), there is a significant increase of the photons which
can ionize OII, NeII, CIII, with decreasing metallicity, the
effect being strongest for those ionic species with ionization
edges closest to the HeII absorption edge. The HeII ionizing
photons are very strongly affected by metallicity (and luminosity)
through the strengths of stellar winds. We also calculated
synthetic spectra and were able to present for the first time
predictions of UV spectra of very massive stars at extremely low
metallicities. From these calculations we learned that the
presence of stellar winds leads to observable broad spectral
line features, which might be used for spectral diagnostics,
should such an extreme stellar population be detected at high
redshift.

We find these first steps very encouraging to proceed with our
calculations towards a number of improvements and extensions in
the future. So far, our stellar parameters have been chosen from
simple scaling relations and not from consistent stellar interior
and evolution calculations. While this was certainly sufficient at
the beginning to find out what the basic effects are, we need to
remove this deficiency in a next step to become more
quantitative. We also have to increase the range of effective
temperatures, since the zero age main sequences of very massive
stars are shifted beyond 60000K for metallicities as low as in
this paper (\citealt{bro01b}, \citealt{tum00}, \citealt{chi00},
\citealt{bar01}). In this way, we will be able
to make improved predictions
about the influence of stellar winds on the evolution of very
massive stars and on the evolution of galaxies through deposition
of matter, radiation, momentum and energy. These improved calculations
should also take into account the effects of changes in the chemical abundance
pattern of metals. So far, we have adopted relative abundances as in the
solar system and have only scaled the total metallicity. However, it is very
likely that an early generation of very massive stars will have an abundance
pattern substantially different from the sun, in particular with regard to
the ratio of $\alpha$ to iron group elements. As has been shown by
\cite{pul00} and \cite{vin01} in the case of normal O-stars, this can have a
significant influence on the stellar wind properties.

{\bf Acknowledgments}

It is a pleasure to thank my former Munich University Observatory
colleagues Adi Pauldrach, Joachim Puls and Uwe Springmann for
their active support during the time, when this work was started.
Volker Bromm and Avi Loeb directed my attention to the important
role of very massive stars in the early universe and motivated
much of the work done here. Special thanks go to Fabio Bresolin
and Roberto Mendez for careful reading of the manuscript and
critical remarks. The detailed and very constructive comments of
the referee are gratefully acknowledged.

\appendix

\section{Appendix. The functions $D_{\alpha}$ and $\Delta v_{reg}$}

Eqs.\,\ref{singcond2}, \ref{p} and \ref{Ccrit} contain the function
$D_{\alpha}$ which follows from the derivative of the first
term of Eq.\,\ref{eom2} with respect to y

\begin{equation}
D_{\alpha} = D_{0} + D_{1} - D_{2}
\label{dalpha}
\end{equation}

with

\begin{equation}
D_{0} = 1 + \fre{\alpha_{0}\alpha_{1}}{\fri{\alpha}} log(t), \qquad \qquad
D_{2} = \fre{\gamma}{\fri{\alpha}} log(\hat{n}),
\label{d0}
\end{equation}

\begin{equation}
D_{1} = \fre{1}{\fri{\alpha}}
\lbrace \fre{\alpha_{0}\alpha_{1}}{\fri{\alpha+1}}
\fre{1}{\fri{ln(10)}} - \fre{h}{\fri{1-h}} +
(\alpha+1)\fre{(1-\lambda)^{\alpha+1}}{\fri{1-(1-\lambda)^{\alpha+1}}}H
\rbrace,
\label{d1}
\end{equation}

and

\begin{equation}
H = u^{2}h +
\fre{\alpha_{0}\alpha_{1}}{\fri{\alpha+1}}(1-\lambda)log(1-\lambda).
\label{H}
\end{equation}

Note that with constant $\alpha$ (i.e. $\alpha_{1}=0$) $D_{\alpha}$
simplifies to

\begin{equation}
D_{\alpha} = D_{\alpha}^{0}, \qquad \qquad
D_{\alpha}^{0} = 1 + \fre{1}{\fri{\alpha}}
\lbrace (\alpha+1)\fre{(1-\lambda)^{\alpha+1}}{\fri{1-(1-\lambda)^{\alpha+1}}}
u^{2}h - \fre{h}{\fri{1-h}}\rbrace
\label{dalpha0}
\end{equation}

describing the influence of the finite cone angle correction factor on
the derivative with respect to y (see \cite{pau86}). Neglecting
the finite cone angle correction factor leads to $D_{\alpha}^{0}=1$.

The function $\Delta v^{2}_{reg}$ in Eq.\,\ref{regcond2} is given
by

\begin{equation}
\Delta v^{2}_{reg} = -\fre{1}{\fri{2a_{0}}} \lbrace a_{1} +
(a_{1}^{2}-4a_{0}a_{2})^{0.5} \rbrace \label{delv}
\end{equation}

where the coefficients $a_{0,1,2}$ are defined as

\begin{equation}
a_{0} = \fre{a}{\fri{1-\alpha_{eff}}} \fre{2}{\fri{u}} D_{u} -
\fre{v_{s}^{2}}{\fri{v_{esc}^{2}}} \fre{4}{\fri{u^{2}}},
\label{a0}
\end{equation}

\begin{equation}
a_{1} = -\alpha_{eff} \lbrace \fre{a}{\fri{1-\alpha_{eff}}} \rbrace ^{2}
v_{esc}^{2} D_{v}, \qquad \qquad
a_{2} = (\alpha_{eff})^{2} \lbrace \fre{a}{\fri{1-\alpha_{eff}}} \rbrace ^{2}
v_{esc}^{2} v_{s}^{2}.
\label{a1}
\end{equation}

The functions $D_{u}$ and $D_{v}$ are connected with the partial derivatives
of the function f with respect to u and v

\begin{equation}
\fre{1}{\fri{f}}\fre{\partial f}{\fri{\partial u}}  =
\fre{2}{\fri{u}} D_{u}, \qquad \qquad
\fre{1}{\fri{f}}\fre{\partial f}{\fri{\partial v^{2}}}  =
\fre{1}{\fri{v^{2}}} D_{v}
\label{Du1}
\end{equation}

and calculated by

\begin{equation}
D_{u}  = \delta D_{W} D_{\delta} - \fre{1-h/2}{\fri{1-h}} D_{\lambda}, \qquad
\qquad
D_{v}  = \fre{h}{\fri{1-h}} D_{\lambda} - \fre{\delta}{\fri{2}} D_{\delta},
\label{Du2}
\end{equation}

where

\begin{equation}
D_{\lambda}  = 1 - (\alpha+1)\lambda
\fre{(1-\lambda)^{\alpha}}{\fri{1-(1-\lambda)^{\alpha+1}}},
\label{Dlambda}
\end{equation}

and

\begin{equation}
D_{\delta}  = 1 + \fre{\delta_{0}\delta_{1}}{\fri{\delta}} log(\hat{n}),
\qquad \qquad
D_{W}  = 1 - \fre{1}{\fri{4}} \fre{u^{2}}{\fri{W(u)(1-u^{2})^{0.5}}}.
\label{Ddelta}
\end{equation}

\begin{deluxetable}{llrrrrrr}
\tablecaption{New force multiplier parameters \label{fmpsmet}}
\tablewidth{0pt}
\tablehead{
\colhead{T$_{eff}$} & \colhead{Z/Z$_{\odot}$} &
\colhead{$\hat{k}$} &
\colhead{$\alpha_{o}$} & \colhead{$\alpha_{1}$} &
\colhead{$\delta_{o}$} &
\colhead{$\delta_{1}$} &
\colhead{$\gamma$}}
\startdata
 40000 & 1.0    & 0.074 & 0.697 &  0.018 &  0.199 &  0.037 &  0.032\\
       & 0.2    & 0.054 & 0.715 &  0.029 &  0.192 & -0.046 &  0.030\\
       & 0.001  & 0.010 & 0.778 &  0.061 &  0.018 & -0.726 & -0.028\\
       & 0.0001 & 0.008 & 0.488 & -0.013 & -0.063 &  0.457 & -0.067\\
 50000 & 1.0    & 0.084 & 0.695 &  0.021 &  0.197 & -0.134 & -0.007\\
       & 0.2    & 0.041 & 0.798 &  0.042 &  0.233 & -0.019 &  0.019\\
       & 0.01   & 0.031 & 0.728 &  0.054 &  0.068 &  0.185 & -0.024\\
       & 0.001  & 0.018 & 0.673 &  0.050 & -0.026 &  0.556 & -0.073\\
       & 0.0001 & 0.008 & 0.548 & -0.003 & -0.084 &  0.462 & -0.105\\
 60000 & 1.0    & 0.028 & 0.668 & -0.008 &  0.266 &  0.249 & -0.065\\
       & 0.2    & 0.023 & 0.743 &  0.027 &  0.223 &  0.198 &  0.045\\
       & 0.01   & 0.026 & 0.722 &  0.062 &  0.163 &  0.096 &  0.004\\
       & 0.001  & 0.016 & 0.662 &  0.062 &  0.036 &  0.714 & -0.064\\
       & 0.0001 & 0.004 & 0.773 &  0.061 & -0.133 &  0.045 & -0.154\\
\enddata
\end{deluxetable}

\clearpage
\newpage

\begin{deluxetable}{llllrrrr}
\tablecaption{Stellar parameter and stellar wind properties \label{windprop}}
\tablewidth{0pt}
\tablehead{
\colhead{log L/L$_{\odot}$} & \colhead{T$_{eff}$ \tablenotemark{a}} &
\colhead{log g \tablenotemark{a}} &
\colhead{R/R$_{\odot}$} &
\colhead{Z/Z$_{\odot}$} &
\colhead{v$_{\infty}$ \tablenotemark{b}} &
\colhead{$\dot{M}$ \tablenotemark{b}} &
\colhead{log D$_{\rm mom}$ \tablenotemark{c}}}
\startdata
 7.03 & 60000. & 3.95 & 30.38 & 1.0    & 1043.4 & 483.6 & 31.24 \\
      &        &      &       & 0.2    & 1176.4 & 138.5 & 30.75 \\
      &        &      &       & 0.01   & 1251.5 &  59.9 & 30.41 \\
      &        &      &       & 0.001  &  890.9 &  19.5 & 29.78 \\
      &        &      &       & 0.0001 &  711.7 &   2.2 & 28.73 \\
      & 50000. & 3.63 & 43.76 & 1.0    & 1211.5 & 591.4 & 31.47 \\
      &        &      &       & 0.2    & 1372.1 & 263.4 & 31.18 \\
      &        &      &       & 0.01   & 1182.8 &  63.9 & 30.50 \\
      &        &      &       & 0.001  &  911.2 &  17.5 & 29.82 \\
      &        &      &       & 0.0001 &  612.7 &   2.4 & 28.80 \\
      & 40000. & 3.25 & 68.32 & 1.0    &  943.0 & 528.2 & 31.41 \\
      &        &      &       & 0.2    &  971.7 & 226.9 & 31.06 \\
      &        &      &       & 0.001  &  730.9 &  10.0 & 29.58 \\
      &        &      &       & 0.0001 &  424.3 &   1.2 & 28.45 \\
 6.91 & 60000. & 3.99 & 26.24 & 1.0    & 1215.0 & 116.8 & 30.66 \\
      &        &      &       & 0.2    & 1399.2 &  41.7 & 30.27 \\
      &        &      &       & 0.01   & 1280.6 &  26.3 & 30.04 \\
      &        &      &       & 0.001  &  958.8 &   4.8 & 29.17 \\
      &        &      &       & 0.0001 &  495.9 &   0.2 & 27.59 \\
      & 50000. & 3.68 & 38.06 & 1.0    & 1830.9 & 219.5 & 31.19 \\
      &        &      &       & 0.2    & 1731.2 & 109.4 & 30.86 \\
      &        &      &       & 0.01   & 1307.2 &  27.9 & 30.15 \\
      &        &      &       & 0.001  & 1035.4 &   5.8 & 29.37 \\
      &        &      &       & 0.0001 &  555.9 &   0.3 & 27.92 \\
      & 40000. & 3.28 & 59.49 & 1.0    & 1176.1 & 211.1 & 31.08 \\
      &        &      &       & 0.2    & 1248.6 &  92.6 & 30.75 \\
      &        &      &       & 0.001  &  879.4 &   3.2 & 29.14 \\
      &        &      &       & 0.0001 &  425.3 &   0.1 & 27.46 \\
 6.76 & 60000. & 4.04 & 22.29 & 1.0    & 1272.9 &  31.9 & 30.08 \\
      &        &      &       & 0.2    & 1622.5 &  16.0 & 29.89 \\
      &        &      &       & 0.01   & 1405.3 &   8.7 & 29.56 \\
      &        &      &       & 0.001  &  939.9 &   0.9 & 28.44 \\
      & 50000. & 3.73 & 32.10 & 1.0    & 2205.2 & 100.6 & 30.90 \\
      &        &      &       & 0.2    & 1839.5 &  55.0 & 30.56 \\
      &        &      &       & 0.01   & 1443.3 &  11.4 & 29.77 \\
      &        &      &       & 0.001  & 1029.8 &   1.9 & 28.85 \\
      & 40000. & 3.34 & 50.12 & 1.0    & 1476.5 &  79.7 & 30.72 \\
      &        &      &       & 0.2    & 1423.6 &  40.9 & 30.41 \\
      &        &      &       & 0.001  &  926.7 &   1.0 & 28.64 \\
 6.57 & 60000. & 4.11 & 17.78 & 1.0    & 1799.7 &   9.2 & 29.64 \\
      &        &      &       & 0.2    & 2147.5 &   5.3 & 29.48 \\
      &        &      &       & 0.01   & 1507.4 &   2.4 & 28.99 \\
      &        &      &       & 0.001  &  745.4 &  0.07 & 27.14 \\
      & 50000. & 3.79 & 25.74 & 1.0    & 2270.1 &  48.3 & 30.54 \\
      &        &      &       & 0.2    & 1947.6 &  24.3 & 30.18 \\
      &        &      &       & 0.01   & 1518.0 &   4.1 & 29.30 \\
      &        &      &       & 0.001  &  923.0 &   0.4 & 28.14 \\
      & 40000. & 3.41 & 40.18 & 1.0    & 1545.0 &  36.3 & 30.35 \\
      &        &      &       & 0.2    & 1641.5 &  16.6 & 30.04 \\
      &        &      &       & 0.001  &  900.2 &   0.2 & 28.01 \\
 6.42 & 60000. & 4.16 & 15.07 & 1.0    & 2426.5 &   4.3 & 29.41 \\
      &        &      &       & 0.2    & 2469.1 &   2.7 & 29.21 \\
      &        &      &       & 0.01   & 1535.8 &   1.0 & 28.58 \\
      & 50000. & 3.85 & 21.71 & 1.0    & 2279.4 &  29.0 & 30.29 \\
      &        &      &       & 0.2    & 2021.0 &  13.3 & 29.90 \\
      &        &      &       & 0.01   & 1525.7 &   2.0 & 28.96 \\
      &        &      &       & 0.001  &  771.2 &  0.13 & 27.47 \\
      & 40000. & 3.46 & 33.93 & 1.0    & 1675.1 &  19.7 & 30.08 \\
      &        &      &       & 0.2    & 1746.5 &   9.1 & 29.77 \\
      &        &      &       & 0.001  &  846.9 &  0.10 & 27.51 \\
 6.30 & 60000. & 4.21 & 13.11 & 1.0    & 2939.8 &   2.6 & 29.24 \\
      &        &      &       & 0.2    & 2693.1 &   1.6 & 29.00 \\
      &        &      &       & 0.01   & 1539.4 &  0.51 & 28.25 \\
      & 50000. & 3.89 & 18.88 & 1.0    & 2301.6 &  18.9 & 30.07 \\
      &        &      &       & 0.2    & 2056.2 &   8.4 & 29.67 \\
      &        &      &       & 0.01   & 1513.5 &   1.1 & 28.68 \\
      & 40000. & 3.50 & 29.51 & 1.0    & 1767.5 &  12.2 & 29.87 \\
      &        &      &       & 0.2    & 1809.7 &   5.7 & 29.55 \\
      &        &      &       & 0.001  &  795.4 &  0.04 & 27.07 \\
\enddata
\tablenotetext{a}{T$_{eff}$ in Kelvin, g in cgs}
\tablenotetext{b}{v$_{\infty}$ in km/sec, $\dot{M}$ in 10$^{-6}$
M$_{\odot}$/yr}
\tablenotetext{c}{D$_{\rm mom}$ in cgs}
\end{deluxetable}

\clearpage
\newpage

\begin{deluxetable}{lllrrr}
\tablecaption{Fit parameters for mass-loss formula \label{mdtab}}
\tablewidth{0pt}
\tablehead{
\colhead{y} & \colhead{T$_{eff}$ \tablenotemark{a}} &
\colhead{$a_{0}$} &
\colhead{$a_{1}$} &
\colhead{$a_{2}$}}
\startdata
$[Z]_{min}$ & 60000. &  -3.40 & -0.40 & -0.65 \\
            & 50000. &  -3.85 & -0.05 & -0.60 \\
        & 40000. &  -4.45 &  0.35 & -0.80 \\
$Q_{min}$   & 60000. &  -8.00 & -1.20 &  2.15 \\
            & 50000. & -10.35 &  3.25 &  0.00 \\
        & 40000. & -11.75 &  3.65 &  0.00 \\
$Q_{0}$     & 60000. &  -5.99 &  1.00 &  1.50 \\
            & 50000. &  -4.85 &  0.50 &  1.00 \\
        & 40000. &  -5.20 &  0.93 &  0.85 \\
\enddata
\tablenotetext{a}{T$_{eff}$ in Kelvin}
\end{deluxetable}

\newpage

\begin{figure}
\begin{center}
\figurenum{4}
\epsscale{1.}
\plotone{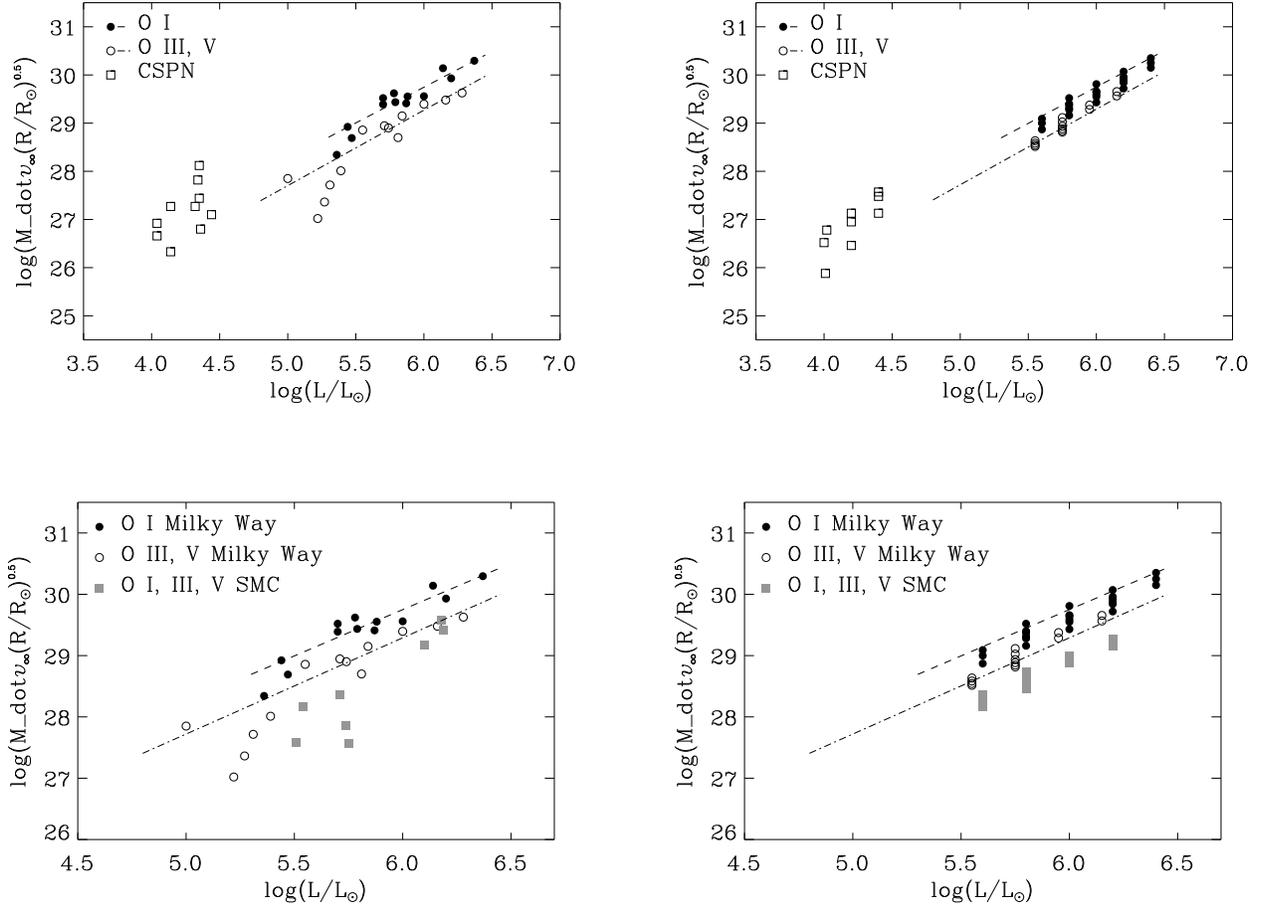}
\caption{Top: Modified wind momenta of galactic O-stars and
Central Stars of PN as a function of luminosity. Left: Observations (the
dashed and dashed-dotted curves are linear regressions for the different
O-star luminosity classes); right: model calculations as described in the
text compared to the observed regression curves.
Bottom: Modified wind momenta of O-stars
in the galaxy and SMC as a function of luminosity. Left: Observations;
right: calculations as described in the text. The linear regressions
for galactic O-stars of the top figure are also shown.
\label{fig4}}
\end{center}
\end{figure}

\begin{figure}
\begin{center}
\figurenum{5}
\epsscale{0.8}
\plotone{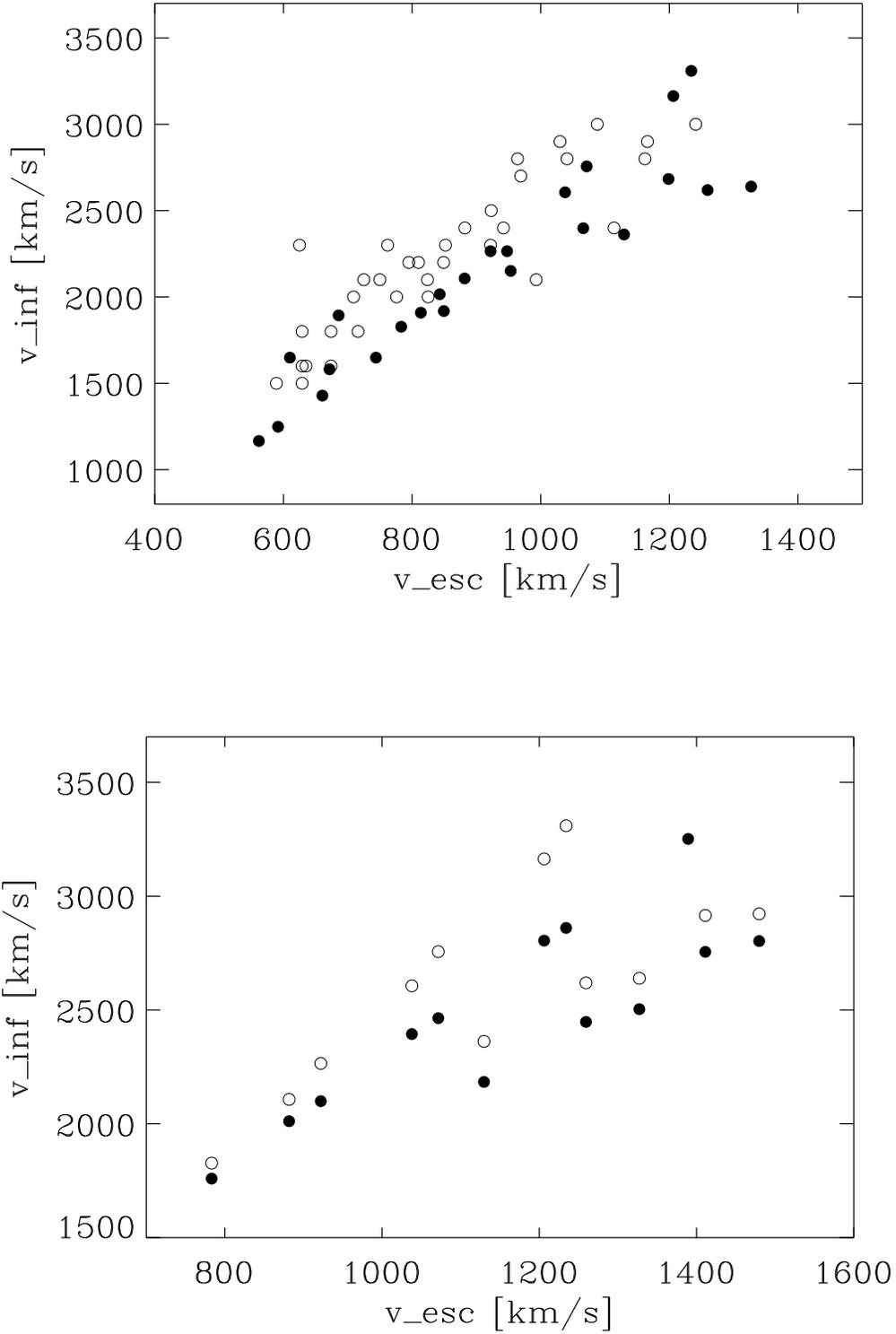}
\caption{Top: Terminal velocity of galactic O-stars versus
photospheric escape velocity. Open circles represent observations taken
from Lamers et al.\/(1995). Solid circles corresponds to the calculations as
described in the text.
Bottom: The influence of metallicity on the computations
of terminal velocities for O main sequence stars. Open circles correspond to
the calculations for galactic metallicity.
Solid circles represent calculations for Z = 0.2 Z$_{\odot}$.
\label{fig5}}
\end{center}
\end{figure}

\begin{figure}
\begin{center}
\figurenum{6}
\epsscale{1.0}
\plotone{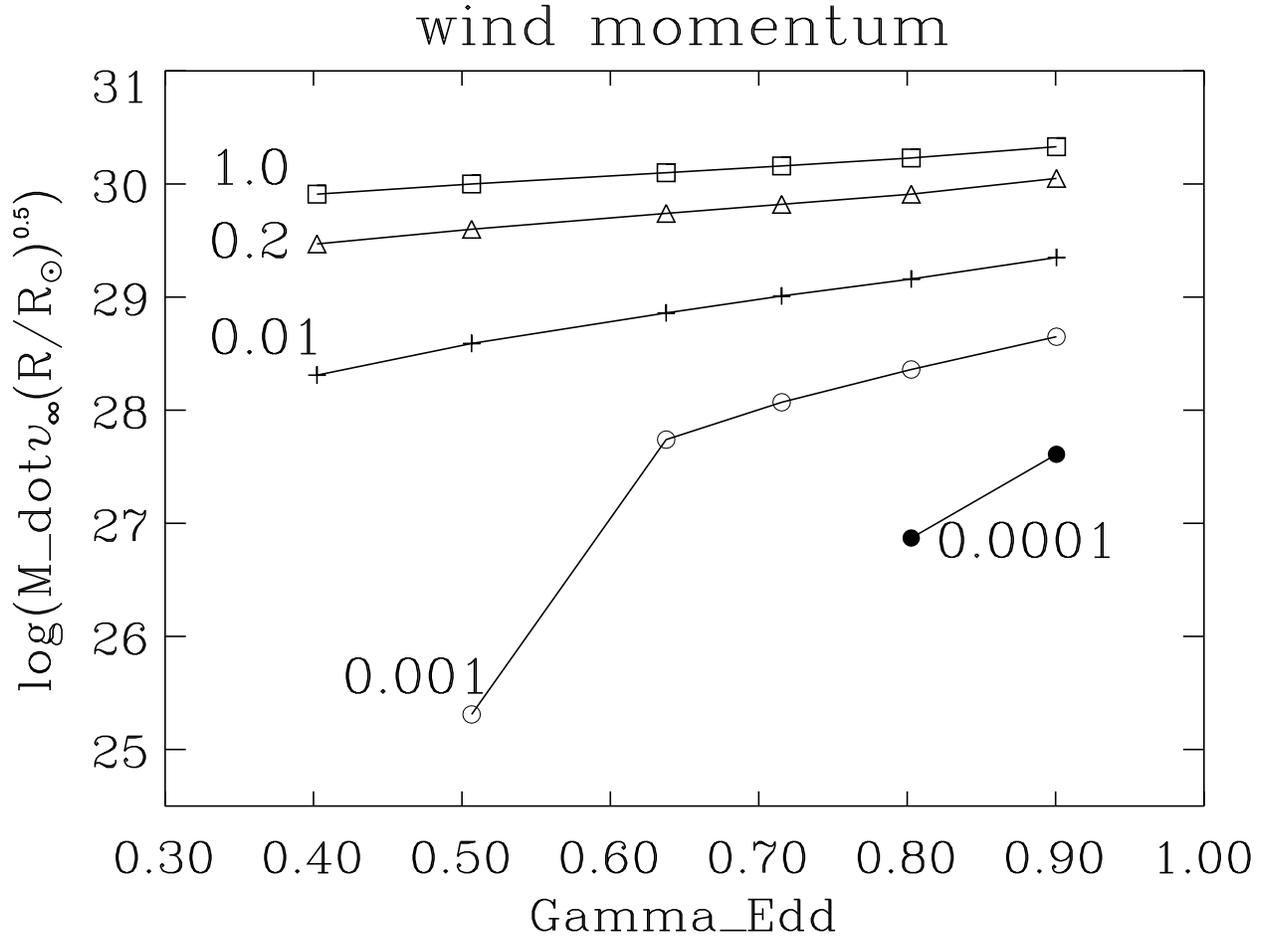}
\caption{Modified stellar wind momenta as a function of
$\Gamma$, the ratio of electron scattering radiative acceleration
to gravitational acceleration. All models have been calculated for
T$_{eff}$ = 50000 K and log L/L$_{\odot}$ = 6.26. The different sequences are
labelled by the value of the metallicity Z/Z$_{\odot}$ adopted.
\label{fig6}}
\end{center}
\end{figure}

\begin{figure}
\begin{center}
\figurenum{7}
\epsscale{1.0}
\plotone{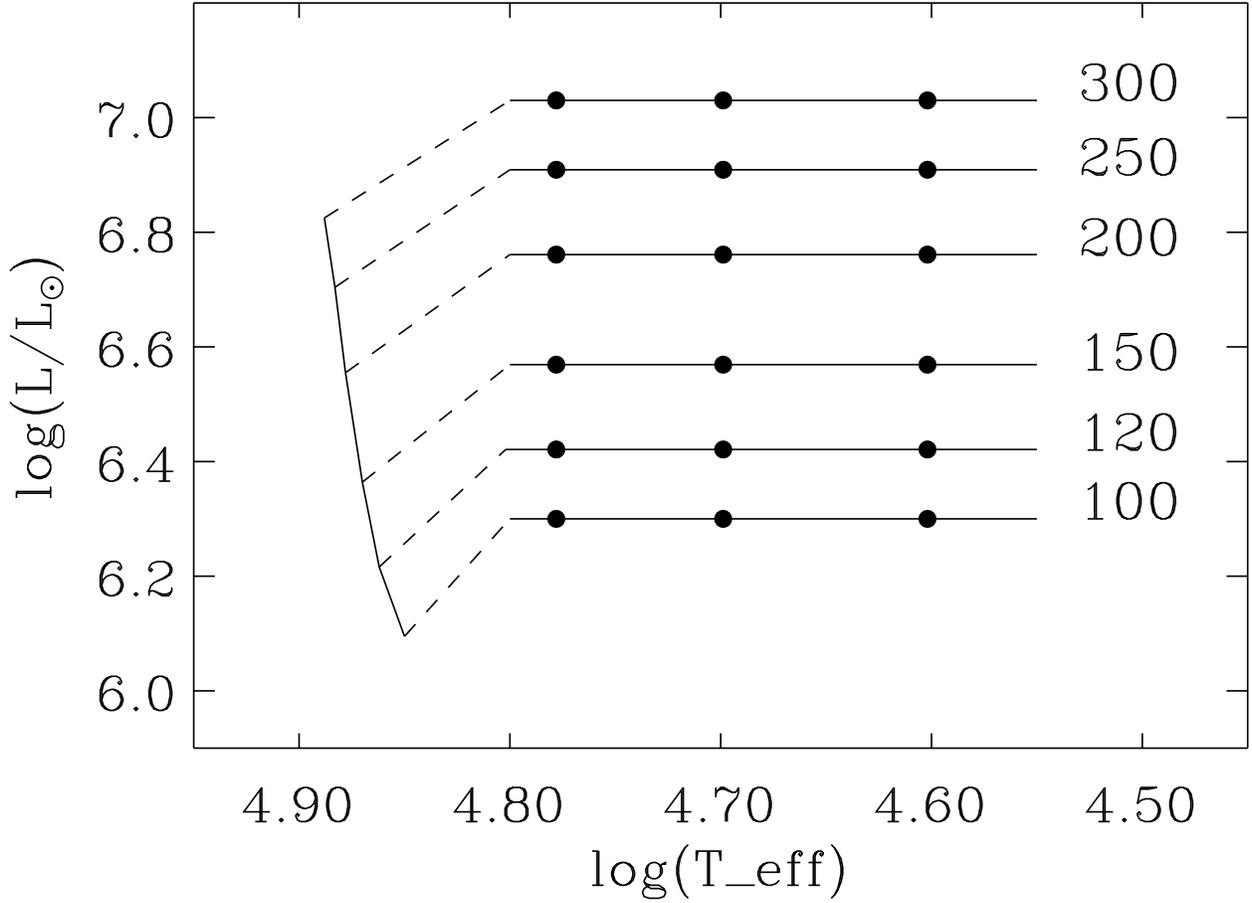}
\caption{The HRD of the simplified stellar models adopted
for the stellar wind calculations. The tracks are labelled by the stellar mass
in solar units. Luminosities and effective temperatures of the wind models
calculated for this study are given by filled circles.
\label{fig7}}
\end{center}
\end{figure}

\begin{figure}
\begin{center}
\figurenum{8}
\epsscale{1.0}
\plotone{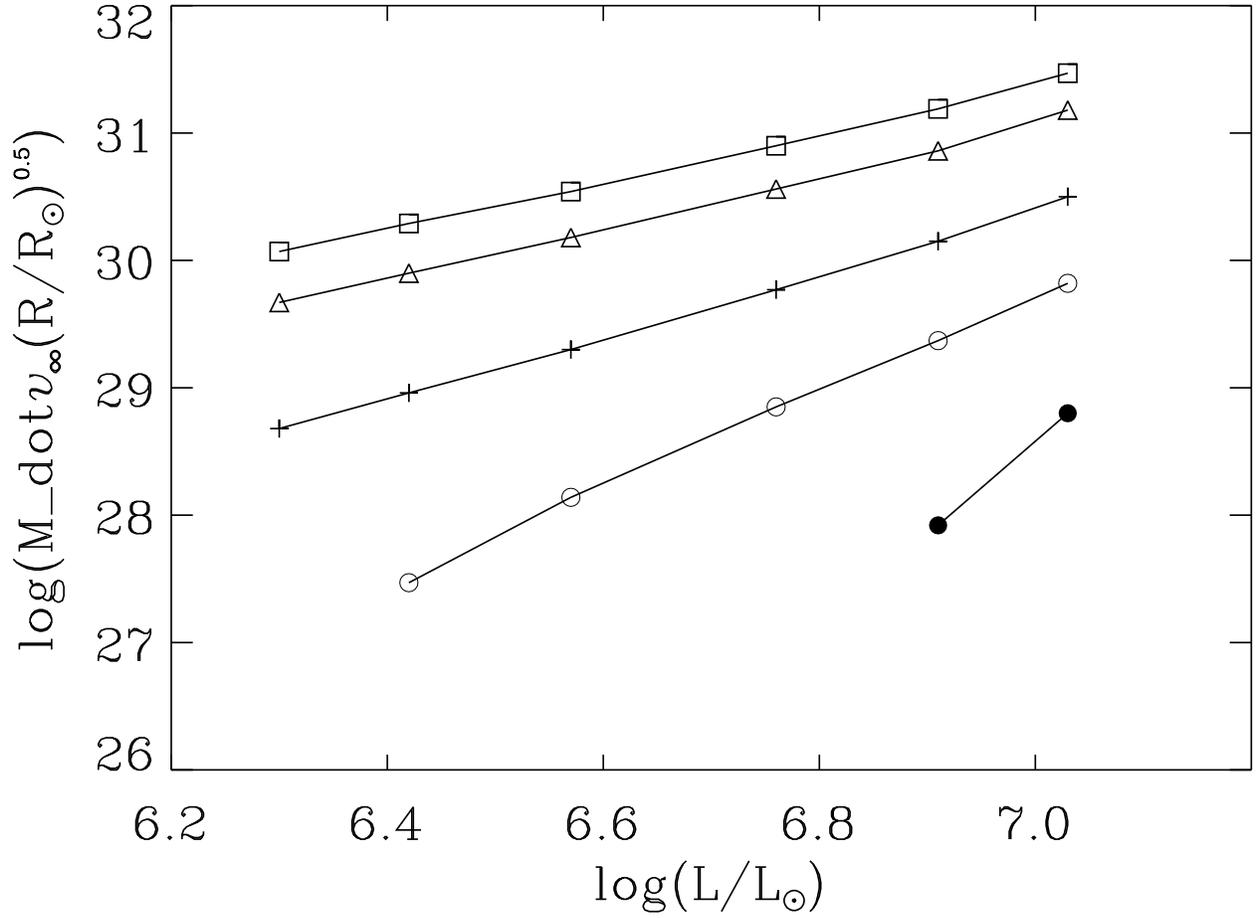}
\caption{Modified stellar wind momentum calculated as a
function of stellar luminosity for models of different metallicity with
Z/Z$_{\odot}$ = 1.0 (squares), 0.2 (triangles), 10$^{-2}$ (plus signs),
10$^{-3}$ (open circles) and 10$^{-4}$ (solid circles).
\label{fig8}}
\end{center}
\end{figure}

\begin{figure}
\begin{center}
\figurenum{9}
\epsscale{1.}
\plotone{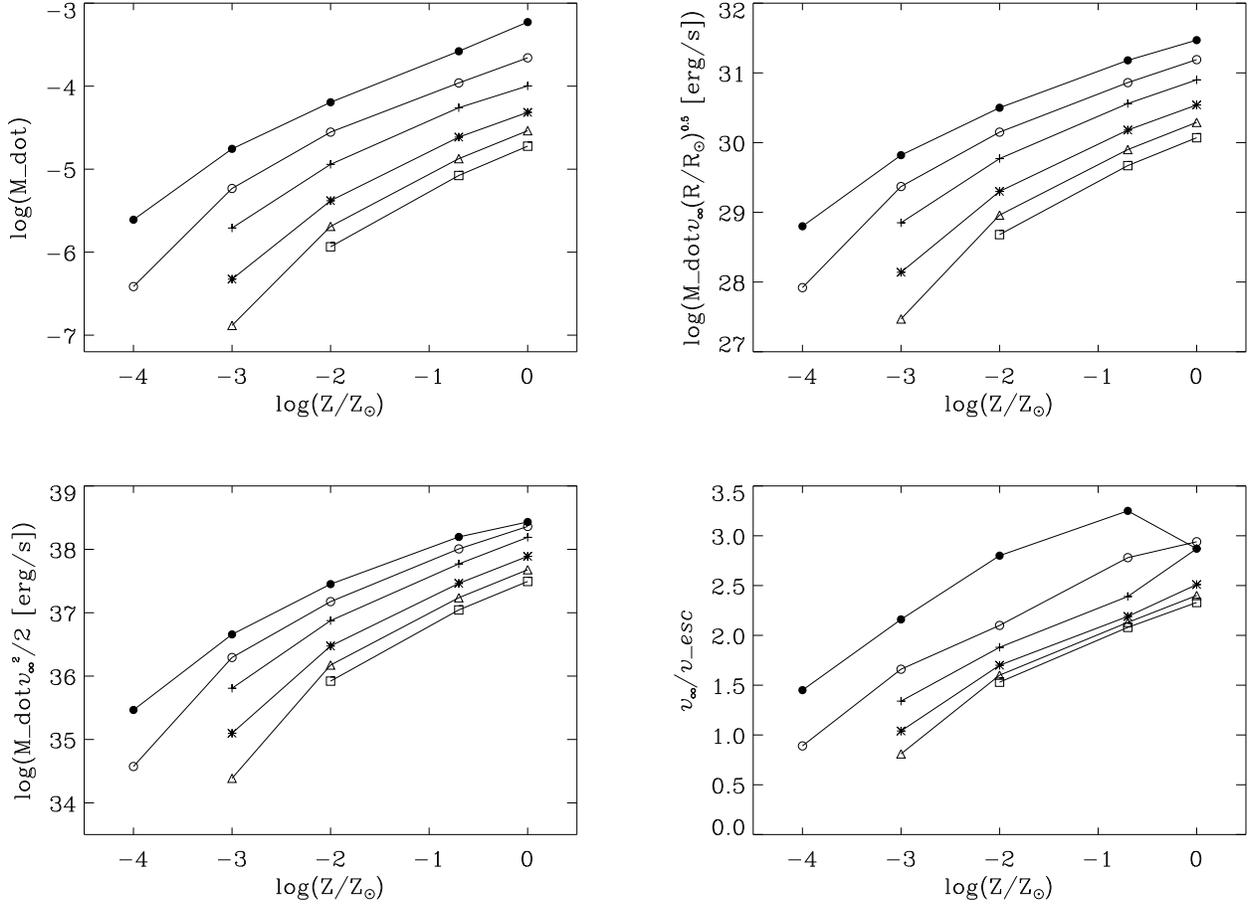}
\caption{Stellar wind properties as a function of metallicity
for models of different luminosity with log L/L$_{\odot}$ = 7.03 (solid
circles), 6.91 (open circles), 6.76 (plus signs), 6.57 (asterisks), 6.42
(triangles), 6.30 (squares). Upper left: mass-loss rates, upper right:
modified stellar wind momentum, lower left: wind energy, lower right: terminal
velocity in units of photospheric escape velocity.
\label{fig9}}
\end{center}
\end{figure}

\begin{figure}
\begin{center}
\figurenum{10}
\epsscale{0.8}
\plotone{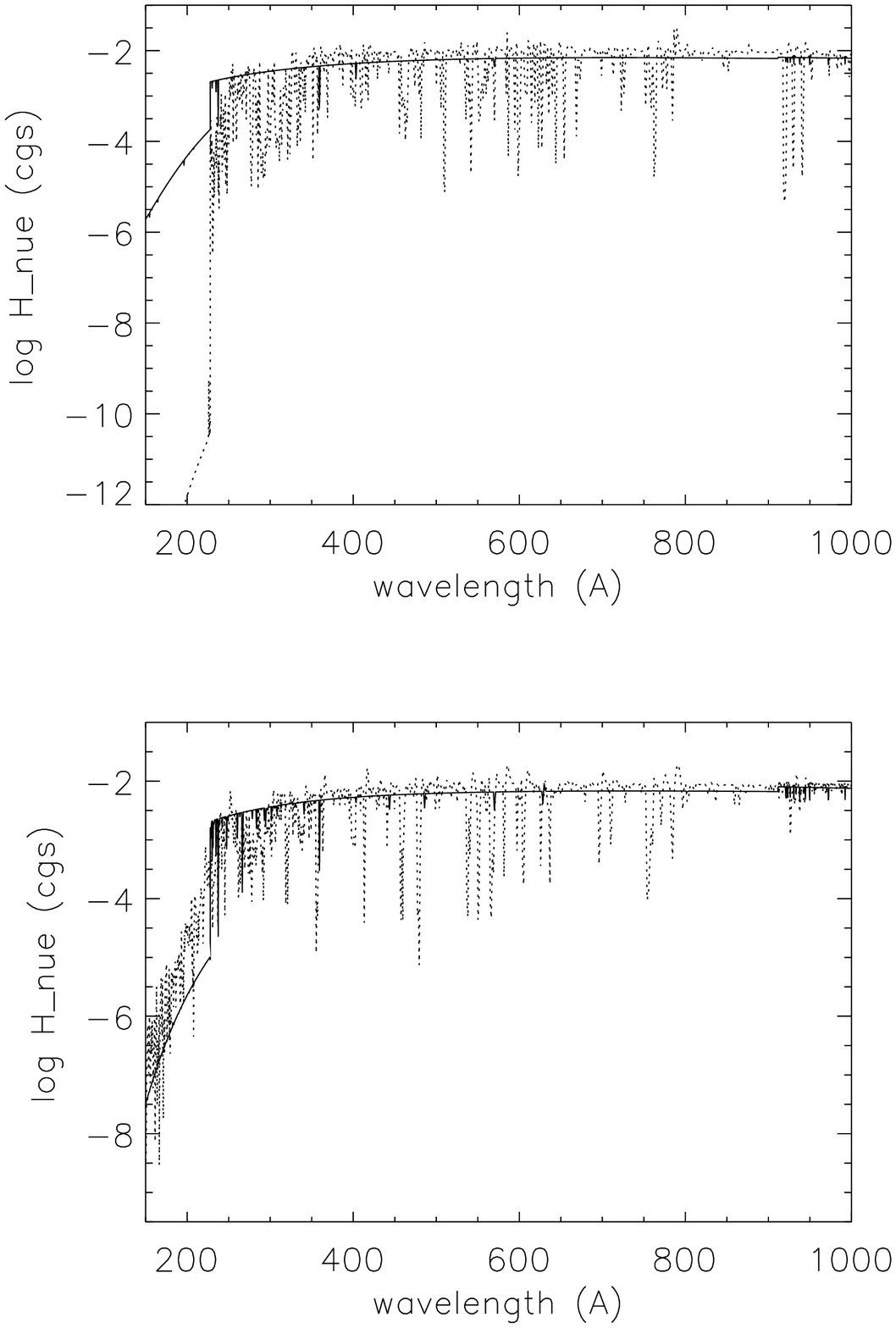}
\caption{Logarithm of emergent EUV/FUV Eddington flux H$_{\nu}$
as a function of wavelength for models with T$_{eff}$ = 60000K and two
metallicities, Z/Z$_{\odot}$ = 1.0 (dotted) and 10$^{-4}$ (solid).
Upper panel: log L/L$_{\odot}$ = 6.91, lower panel: log L/L$_{\odot}$ = 6.57.
Note the dramatic differences in the HeII continuum shortward of 227 A.
\label{fig10}}
\end{center}
\end{figure}

\begin{figure}
\begin{center}
\figurenum{11}
\epsscale{1.0}
\plotone{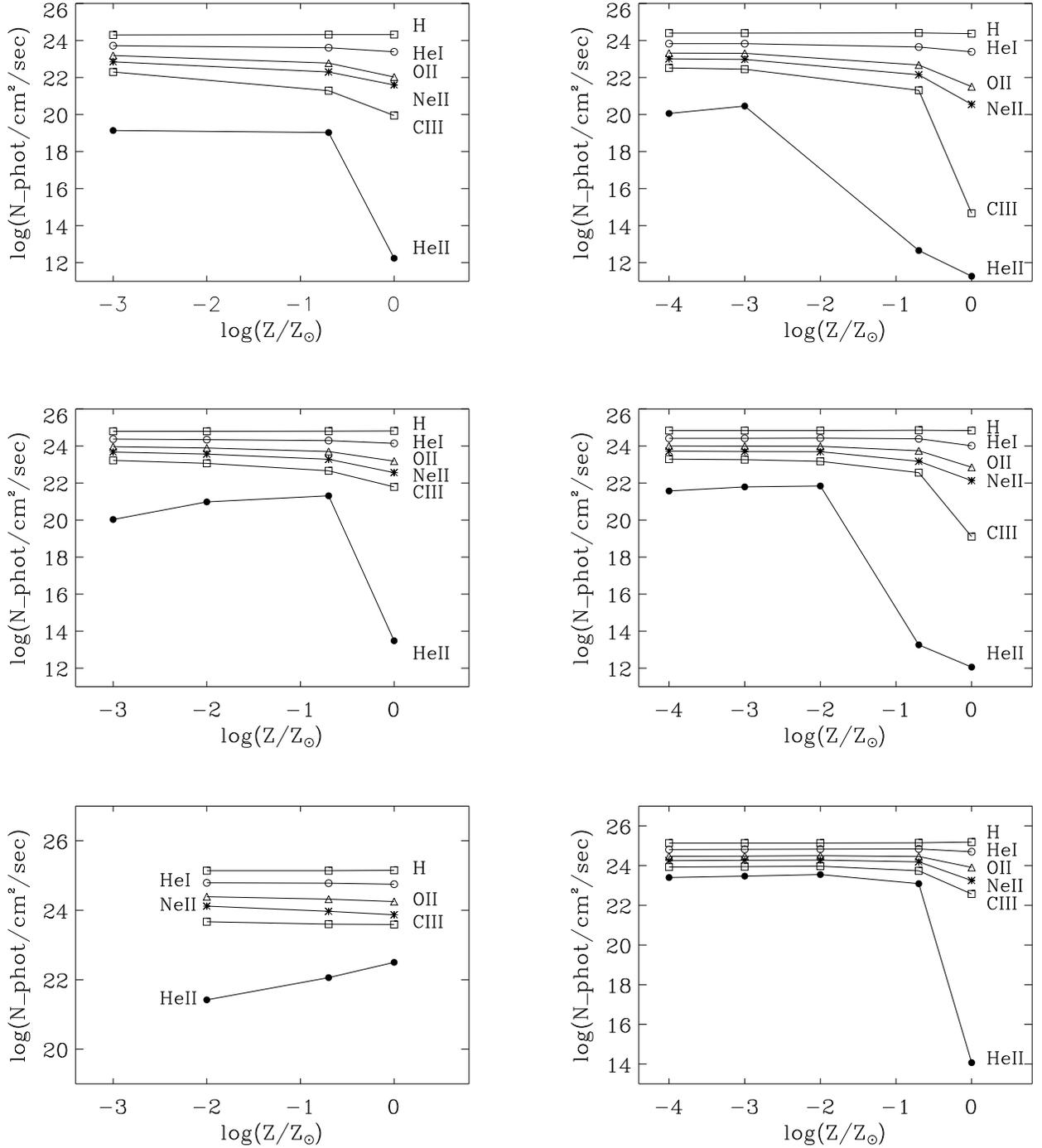}
\caption{Logarithm of the emitted number of ionizing photons as a
function of stellar metallicity (see text). Left: log L/L$_{\odot}$ = 6.42;
right: log L/L$_{\odot}$ = 7.03. Top panels: T$_{eff}$ = 40000K;
middle panels: T$_{eff}$ = 50000K; bottom panels: T$_{eff}$ = 60000K.
\label{fig11}}
\end{center}
\end{figure}

\begin{figure}
\begin{center}
\figurenum{12}
\epsscale{1.0}
\plotone{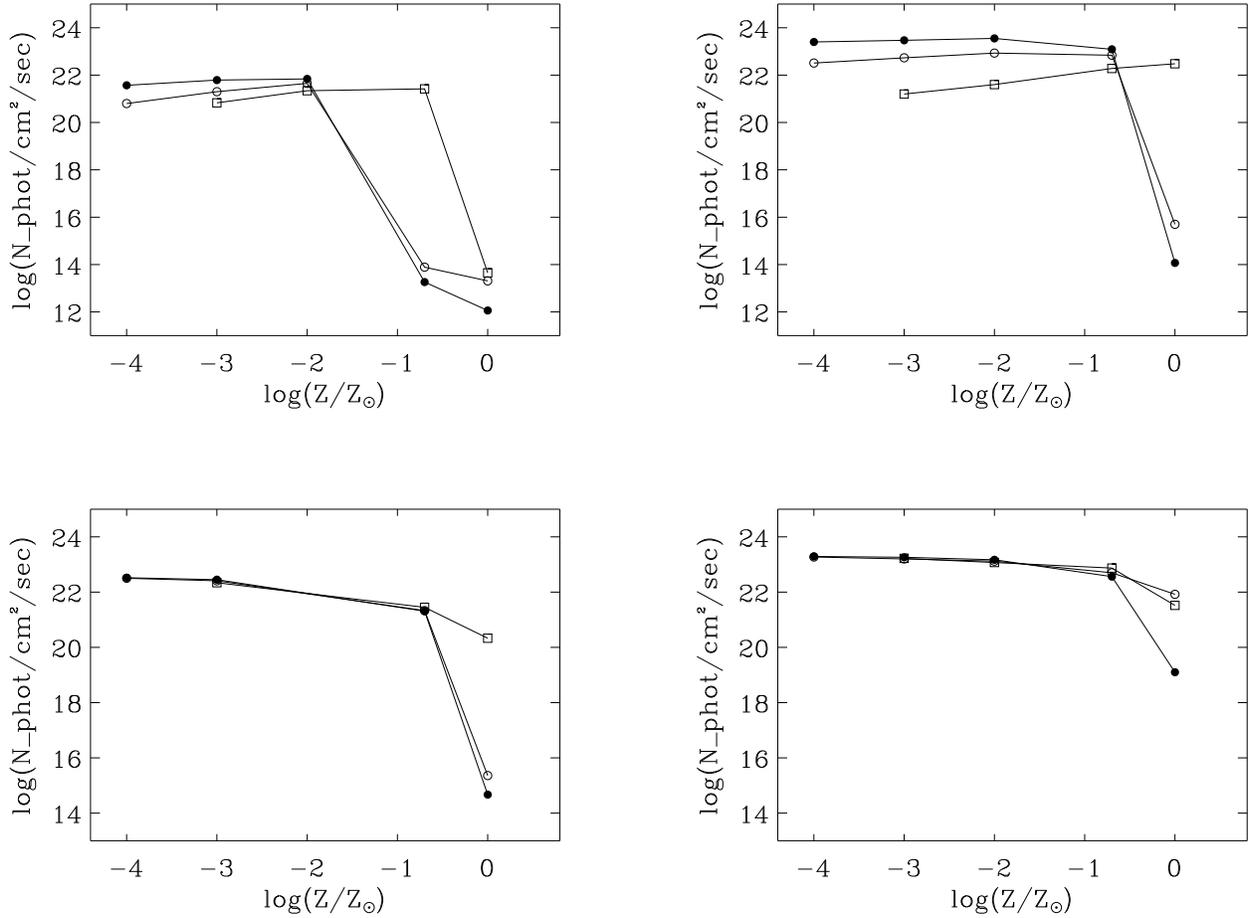}
\caption{Logarithm of the emitted number of ionizing photons as a
function of stellar metallicity (see text). Upper left panel: HeII-photons
at T$_{eff}$ = 50000K, log L/L$_{\odot}$ = 6.76 (squares), 6.91 (circles),
7.03 (solid circles). Upper right panel: HeII-photons
at T$_{eff}$ = 60000K, log L/L$_{\odot}$ = 6.57 (squares), 6.91 (circles),
7.03 (solid circles). Lower panels: CIII-photons at T$_{eff}$ = 40000K (left)
and 50000K (right) and luminosities log L/L$_{\odot}$ = 6.57 (squares),
6.91 (circles), 7.03 (solid circles).
\label{fig12}}
\end{center}
\end{figure}

\begin{figure}
\begin{center}
\figurenum{13}
\epsscale{0.75}
\plotone{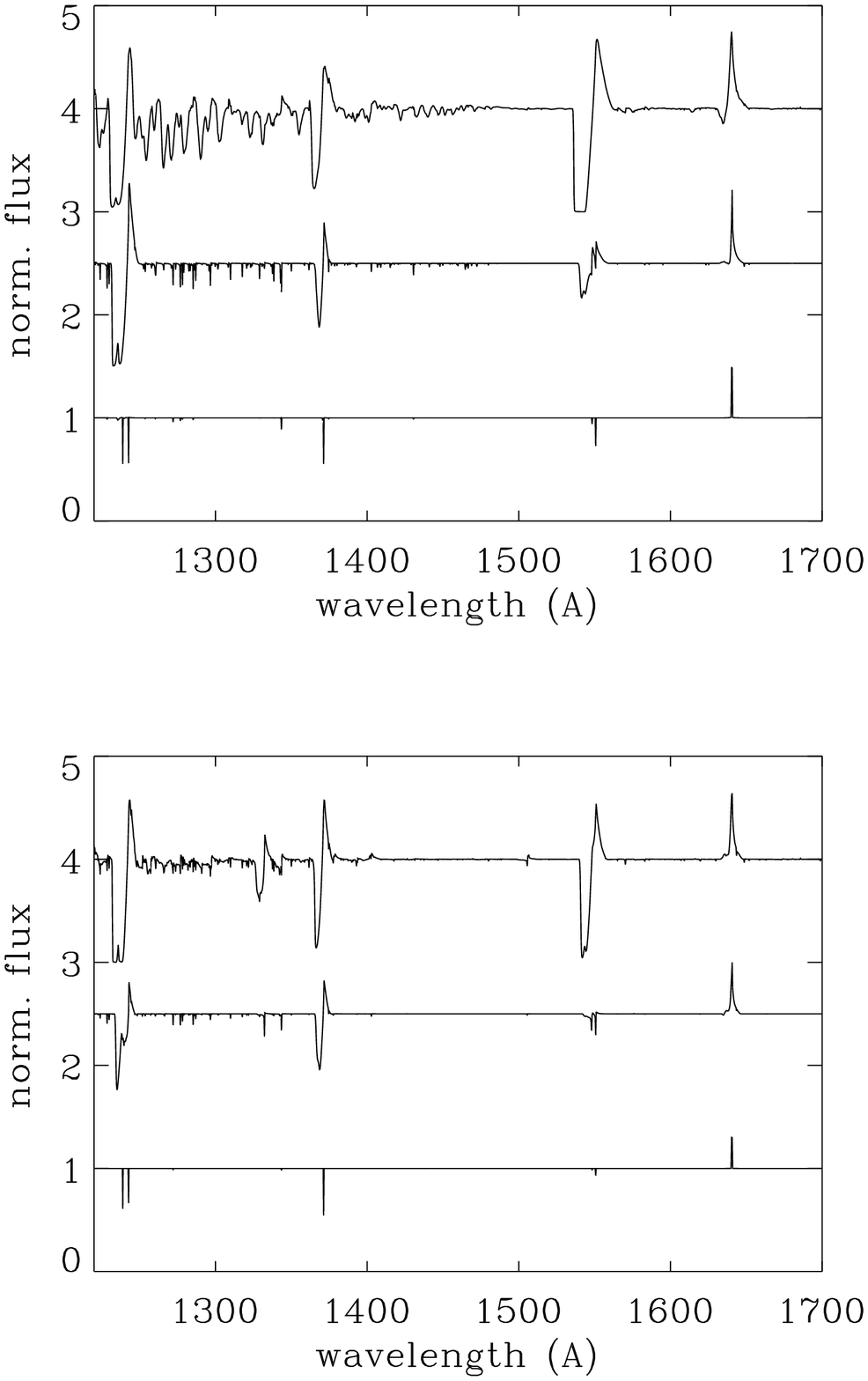}
\caption{Continuum rectified UV-spectra longward of Lyman-alpha
for models with log L/L$_{\odot}$ = 6.91 and T$_{eff}$ = 50000K (top panel)
and 60000K (bottom panel). Each panel displays three metallicities,
Z/Z$_{\odot}$ = 0.2 (top), 10$^{-2}$ (middle), 10$^{-4}$ (bottom). The
strongest line features, which remain present at the lowest metallicity, are
NV $\lambda$ 1250, OV $\lambda$ 1371A, CIV $\lambda$ 1550 and HeII $\lambda$
1640A.
\label{fig13}}
\end{center}
\end{figure}

\end{document}